%% file: main.tex
\documentclass[12pt]{aastex}

\slugcomment{}

\shorttitle{Intensity-Coupled-Polarization in Instruments with a Continuously Rotating Half-Wave Plate}
\include{document_commands}

\begin{document}

%% LaTeX will automatically break titles if they run longer than
%% one line. However, you may use \\ to force a line break if
%% you desire.

\title{Intensity-Coupled-Polarization in Instruments with a Continuously Rotating Half-Wave Plate}

%% Use \author, \affil, and the \and command to format
%% author and affiliation information.
%% Note that \email has replaced the old \authoremail command
%% from AASTeX v4.0. You can use \email to mark an email address
%% anywhere in the paper, not just in the front matter.
%% As in the title, use \\ to force line breaks.

\input{./tex/authors.tex}
\input{./tex/abstract.tex}

%% Keywords should appear after the \end{abstract} command. The uncommented
%% example has been keyed in ApJ style. See the instructions to authors
%% for the journal to which you are submitting your paper to determine
%% what keyword punctuation is appropriate.

\tableofcontents

\maketitle

\acresetall

\input{./tex/introduction.tex}
%\input{./tex/instrument.tex}
\input{./tex/data_model.tex}

\input{./tex/excess_polarization_maps.tex}
\input{./tex/icp_models.tex}
\input{./tex/hwp_template.tex}

\input{./tex/characterization.tex}
\input{./tex/icp_removal.tex}

%\input{./tex/conclusion.tex}

\input{./tex/acronyms}

\bibliographystyle{aasjournal}
\bibliography{main}{}

\appendix

\input{./tex/appendix_coordinates}

\input{./tex/appendix_dielectric}

\input{./tex/appendix_nl_response}

\end{document}

%% file: document_commands.tex
\usepackage{siunitx}
\usepackage{amsmath}
\usepackage{xcolor}
\usepackage{subfigure}
\usepackage{amsmath}
\usepackage{cuted}
\usepackage{esdiff}
\usepackage{hyperref}
\usepackage[nolist]{acronym}

\newcommand{\comred}[1]{\textcolor{red}{#1}}

%% file: tex/authors.tex
\author{
Joy~Didier\altaffilmark{1, 15*, A} 
Amber~D.~Miller\altaffilmark{1, 15*}
Derek~Araujo\altaffilmark{1} 
Fran\c{c}ois~Aubin\altaffilmark{6, 2*} 
Christopher~Geach\altaffilmark{2} 
Bradley~Johnson\altaffilmark{1} 
Andrei~Korotkov\altaffilmark{9} 
Kate~Raach\altaffilmark{2} 
Benjamin~Westbrook\altaffilmark{12} 
Karl~Young\altaffilmark{2} 
Asad~M.~Aboobaker\altaffilmark{2, 4*} 
Peter~Ade\altaffilmark{3} 
Carlo~Baccigalupi\altaffilmark{5} 
Chaoyun~Bao\altaffilmark{2} 
Daniel~Chapman\altaffilmark{1}  
Matt~Dobbs\altaffilmark{6,7} 
Will~Grainger\altaffilmark{8} 
Shaul~Hanany\altaffilmark{2} 
Kyle~Helson\altaffilmark{9, 16*} 
Seth~Hillbrand\altaffilmark{1} 
Johannes~Hubmayr\altaffilmark{10} 
Andrew~Jaffe\altaffilmark{11} 
Terry~J.~Jones\altaffilmark{2} 
Jeff~Klein\altaffilmark{2} 
Adrian~Lee\altaffilmark{12} 
Michele~Limon\altaffilmark{1, 17*}  
Kevin~MacDermid\altaffilmark{6} 
Michael~Milligan\altaffilmark{2} 
Enzo~Pascale\altaffilmark{18} 
Britt~Reichborn-Kjennerud\altaffilmark{1} 
Ilan~Sagiv\altaffilmark{13}
Carole~Tucker\altaffilmark{3} 
Gregory~S.~Tucker\altaffilmark{9} 
Kyle~Zilic\altaffilmark{2}
}

%% Notice that each of these authors has alternate affiliations, which
%% are identified by the \altaffilmark after each name.  Specify alternate
%% affiliation information with \altaffiltext, with one command per each
%% affiliation.

\altaffiltext{1}{Physics Department, Columbia University, New York, NY 10027} %1 
\altaffiltext{2}{School of Physics and Astronomy and Minnesota Institute for Astrophysics, University of Minnesota/Twin Cities, Minneapolis, MN 55455} %4
\altaffiltext{3}{School of Physics and Astronomy, Cardiff University, Cardiff, CF24 3AA, United Kingdom} %3
\altaffiltext{4}{Jet Propulsion Laboratory, California Institute of Technology, Pasadena, CA 91109} %2
\altaffiltext{5}{Astrophysics Sector, SISSA, Trieste, 34014, Italy} %5
\altaffiltext{6}{Department of Physics, McGill University, Montr´eal, Quebec, H3A 2T8, Canada} %6
\altaffiltext{7}{Canadian Institute for Advanced Research, Toronto, ON, M5G1Z8, Canada} %7
\altaffiltext{8}{Rutherford Appleton Lab, Harwell Oxford, OX11 0QX} %8
\altaffiltext{9}{Department of Physics, Brown University, Providence, RI 02912} %9
\altaffiltext{10}{National Institute of Standards and Technology, Boulder, CO 80305} %10
\altaffiltext{11}{Department of Physics, Imperial College, London, SW7 2AZ, United Kingdom} % 11
\altaffiltext{12}{Department of Physics, University of California, Berkeley, Berkeley, CA 94720} %12
\altaffiltext{13}{Faculty of Physics, Weizmann Institute of Science, Rehovot 76100, Israel} %13
\altaffiltext{15}{Department of Physics and Astronomy, University of Southern California, Los Angeles, CA 90089} %15
\altaffiltext{16}{NASA Goddard Space Flight Center, Greenbelt, MD 20771} %16
\altaffiltext{17}{Department of Physics \& Astronomy, University of Pennsylvania, Philadelphia, PA 19104} %17
\altaffiltext{18}{Department of Physics, La Sapienza Università di Roma, Roma, Italy} %17
\altaffiltext{*}{Current affiliation} %*
\altaffiltext{A}{Corresponding Author: Joy Didier (didier.joy@gmail.com)}

%% file: tex/abstract.tex
\begin{abstract}
We discuss a systematic effect associated with measuring polarization with a continuously rotating half-wave plate. 
The effect was identified with the data from the E and B Experiment (EBEX),  which was a balloon-borne instrument designed to measure the polarization of the CMB as well as that from Galactic dust. The data show polarization fraction larger than 10\% while less than 3\% were expected from instrumental polarization. 
We give evidence that the excess polarization is due to detector non-linearity in the presence of a continuously rotating HWP. 
The non-linearity couples intensity signals into polarization. 
We develop a map-based method to remove the excess polarization. 
Applying this method for the 150 (250) GHz bands data we find that 81\% (92\%)  of the excess polarization was removed. 
Characterization and mitigation of this effect is important for future experiments aiming to measure the CMB B-modes with a continuously rotating HWP.

%We discuss a systematic effect associated with measuring the polarization of the \ac{CMB} with a continuously rotating \ac{HWP} using data from the \ac{EBEX}. 
%\ac{EBEX} was a balloon-borne telescope designed to measure the polarization of the \ac{CMB} as well as that from Galactic dust. 
%We show that \ac{EBEX} measured excess polarization coupled to intensity, and provide a model for the origin of this polarization involving detector non-linearity
%in the presence of a continuously rotating \ac{HWP}.
%We provide a map-based method to remove the excess polarization, showing 81 (92) \% of the excess polarization was removed on the \ac{CMB} in the 150 (250) GHz datasets.
%Characterization and mitigation of this effect is important for future experiments aiming to measure the CMB B-modes on large angular scales using a \ac{HWP}.
\end{abstract}

%% file: tex/introduction.tex
\section{Introduction}

% (A)
Measurements of the \ac{CMB} temperature and polarization provide a 
window on the physical mechanisms that govern the evolution of the Universe.  
%Recent research has focused on 
%measuring two distinct patterns of polarization signal in the \ac{CMB}: curl-free E-modes and weaker, 
%divergence-free B-modes \citep{zaldarriaga_1997}. Gravitational lensing of E-modes by large-scale structures 
%generates B-modes at small angular scales, while inflationary expansion of the Universe near the Big Bang is 
%predicted to leave another, as-yet undetected B-mode signature at intermediate and large angular scales \citep{baumann_2009}. 
%
% (B)
The \ac{EBEX} was a balloon-borne telescope designed to measure the polarization of the \ac{CMB} while simultaneously measuring Galactic foreground emission. 
\ac{EBEX} achieved polarimetry via a stationary wire-grid polarizer and a continuously rotating achromatic \ac{HWP}. 
The use of a continuously rotating \ac{HWP} to modulate incident polarized radiation is a well-known polarimetric technique (see, e.g., \cite{maxipol_hwp_2007,abs_hwp_2014}).  
% for separating polarized sky emission from unpolarized foreground signal
Continuous modulation of polarized signals is useful for mitigating systematic errors in two ways.  
It reduces the impact of low frequency noise in the detectors by moving the 
polarization signal of interest to a higher frequency band, where the detector noise is primarily white.  
In addition, it enables measurement of both the $Q$ and $U$ Stokes polarization parameters without 
differencing polarization sensitive detectors.
%signals from detectors that are sensitive to orthogonal linear polarization states.  
Differencing of signals among detector pairs requires the responsivity and noise to be stable and well characterized, 
while mismatching of the detector beams is a source of systematic error such as intensity to polarization signal coupling
\citep{shimon_2008, bicep2015_systematics}. 

% (C)
A common concern for experiments measuring B-modes, which is a curl pattern in the polarization of the CMB, is intensity coupling to the polarization signal, that we refer to as \ac{ICP}.
Intensity signals from the CMB and from foreground sources (including the atmosphere for ground based experiments)
can be orders of magnitude larger than CMB polarization signals. 
Even low levels of \ac{ICP} add systematic bias to the polarization signal and can induce low frequency noise if the intensity is time variable. 
A common source of \ac{ICP} is \ac{IP}.
Here IP is used in the traditional sense referring to the conversion of intensity to polarization through differential transmission or reflection in optical elements.
Another common source of \ac{ICP} is beam and responsivity mismatch between detector pairs. 
Using a continuously rotating \ac{HWP} can mitigate these sources of \ac{ICP}: the IP is reduced because only optical elements sky-side of the
\ac{HWP} contribute to it, and the pair differencing effects are avoided because polarization is measured without differencing detector pairs \citep{abs_hwp_2014,abs_hwp_2016,polarbear_hwp_2017}. 

The subject of this paper is the analysis of a new mechanism for creating \ac{ICP} in EBEX, generated 
by detector non-linearity in the presence of a rotation synchronous signal generated by a HWP.
%The subject of this paper is the discovery of a new mechanism creating \ac{ICP} that was observed in \ac{EBEX} and affects experiments with a continuously rotating \ac{HWP}, coming
%from detector non-linearity in the presence of the rotation synchronous signal generated by a \ac{HWP}. 
A similar effect has been reported in \cite{polarbear_hwp_2017}. 
Understanding and mitigating this effect will be important for future \ac{CMB} missions using a continuously rotating \ac{HWP}.
This paper describes the excess intensity coupled polarization (\ac{ICP}) observed in \ac{EBEX} maps, outlines two possible sources for 
the excess polarization (\ac{IP} and detector non-linearity), uses data to distinguish between those two origins, and details the method we developed to characterize 
and remove the excess polarization.
Because the magnitude of this ICP is correlated with a rotation synchronous signal generated by the \ac{HWP}, we also discuss sources of this rotation synchronous signal.

The paper is organized as follows.  
In Section 2 we outline the data model of an experiment with a continuously rotating \ac{HWP}.  
\ac{EBEX} maps showing excess polarization are shown in Section 3.  
In Section 4 we provide two models for the physical origins of the \ac{ICP}.
In Section 5 we describe in detail the physical origins of the \ac{HWP} synchronous signal.
%In Section 5 we characterize the \ac{HWP} synchronous signal and describe its physical origin.  
%Section 6 shows the excess polarization observed in sky maps of EBEX data.  
In Section 6 we characterize the \ac{ICP} for each EBEX detector and show with this measurement that we can distinguish between various \ac{ICP} mechanisms.
%We use the estimated leakage parameters for the detectors and the behavior of the \ac{HWP} template 
%to demonstrate that the observed $I \rightarrow P$ leakage is caused primarily by non-linear detector 
%responsivity induced by the \ac{HWP} template. 
In Section 7 we present the method we developed to remove the \ac{ICP} and evaluate its performance on real and simulated data.
%We summarize our results in Section 8. 

%% file: tex/data_model.tex
\section{Data model}

The instrument is modelled by an achromatic \ac{HWP} and a wire grid analyzer. The \ac{HWP} is rotating at a constant speed (in EBEX the rotational frequency was 1.235~\si{\hertz}) and we call $\gamma_t$ the angle between the \ac{HWP} extraordinary axis and the polarizing grid, where the subscript $t$ is used to indicate time-dependence. For a given Stokes vector $\vec{S}^{in}_t$ incident on the receiver, the output Stokes vector $\vec{S}^{out}_t$ at the detectors (integrated over the detector bandwidth) is

%\footnotesize
%\[\arraycolsep=1.4pt\def\arraystretch{2.2}
\arraycolsep=5pt
\begin{align}
\vec{S}^{out}_t &= M_{instr} \, \vec{S}^{in}_t \label{eq: basic mueller matrix} \\ 
              &= M_{lp} \, M_{HWP}(\gamma_t) 
    \left( \begin{array}{c}
      I^{in}_t \\
      Q^{in}_t \\
      U^{in}_t \\
      0 \\
    \end{array} \right) \nonumber \\
M_{instr} = M_{lp} \, M_{HWP}(\gamma_t) &= 
    \frac{1}{2} \, \left(
    \begin{array}{cccc}
    1  & \epsilon \cos(4 \gamma_t \! - \! \Phi_t) &  \epsilon \sin(4 \gamma_t \! - \! \Phi_t) &  0          \\
    1  & \epsilon \cos(4 \gamma_t \! - \! \Phi_t) &  \epsilon \sin(4 \gamma_t \! - \! \Phi_t) &  0 \nonumber\\
    0  & 0                                  &  0                                  &  0 \nonumber\\
    0  & 0                                  &  0                                  &  0 \nonumber\\
    \end{array}
    \right)
\end{align}
%\]
\normalsize

\noindent where $M_{lp}$ and $M_{HWP}(\gamma_t)$ are the Mueller matrices of a linear polarizer and a \ac{HWP}, respectively, $\epsilon$ is the \ac{HWP} polarization modulation efficiency, and $\Phi_t$ is an angle encoding the offset between the sky-fixed $Q$ \& $U$ reference frame and the polarizing grid, as well as the frequency dependent phase delay introduced by the achromatic \ac{HWP} \citep{maxipol_hwp_2007, tomo_thesis}. Details on the coordinates and the instrument and sky frames used throughout the paper are available in Appendix~\ref{appendix: coordinates}.
The detectors are only sensitive to power $I^{out}_t$ computed from Equation~\ref{eq: basic mueller matrix}, and their time-stream $D_t$ is:

%\footnotesize
\begin{align}
D_t &= I^{out}_t  \label{eq: basic detector response} \\
    &=\frac{1}{2} \, \Big (I^{in}_t + \epsilon Q^{in}_t \cos(4\gamma_t - \Phi_t) + \epsilon U^{in}_t \sin(4\gamma_t - \Phi_t) \Big ) \nonumber 
\end{align}
\normalsize

\noindent We separate the incoming Stokes vector $\vec{S}^{in}_t$ into the desired sky signal $\vec{S}^{sky}_t$ and $\vec{S}^{instr}_t$ which corresponds to 
spurious unpolarized and polarized signals from the instrument, giving:
%$\vec{S}^{instr}$ is stationary or slowly varying with time compared to $\vec{S}^{sky}_t$. 
%In a linear approach, we re-write the data model $D_t$ accordingly:

%\footnotesize
\begin{align}
D_t = & \, \frac{1}{2} \, \Big (I^{sky}_t + \epsilon Q^{sky}_t \cos(4\gamma_t - \Phi_t) + \epsilon U^{sky}_t \sin(4\gamma_t - \Phi_t) \Big ) \nonumber\\ 
     &+ A(\gamma_t) + n_t  \label{eq: nominal detector model}
\end{align}
\normalsize

\noindent where $n_t$ is the detector noise and $A(\gamma_t)$ groups spurious instrument signals $\vec{S}^{instr}$ modulated by the \ac{HWP}.
%, including polarized instrument emissions and unpolarized instrument emissions subsequently polarized through \ac{IP}. 
The spurious modulation signal, called from hereafter the \ac{HWP} Synchronous Signal, or HWPSS, is modeled by:

\begin{equation}
A(\gamma_t) = \sum\limits_{j=0}^{\infty} \quad \underbrace{A_{j} \cos(j \gamma_t - 2 \alpha_{j})}_\text{stationary} \quad + 
              \quad  \underbrace{A'_{j} I^{sky}_t \cos(j \gamma_t - 2\alpha'_{j})}_\text{scan modulated} \label{eq: hwp model}
\end{equation}

\noindent where we have distinguished between stationary HWPSS emitted by the instrument and scan modulated HWPSS coupled to $I_t^{sky}$.
All spurious effects are lumped into the $A_{j}$, $A'_j$, $\alpha_{j}$ and $\alpha'_j$ parameters and are phenomenologically allowed to be present at all harmonics $j$. 
To account for time-dependant temperature fluctuations in the instrument, the parameters are allowed to vary linearly with time. 
%on periods slow compared to the \ac{HWP} rotation rate. 
%In practice the first few terms in the expansion are dominant, particularly the 4th harmonic.
%resulting from $Q^{instr}$ and $U^{instr}$ in Equation~\ref{eq: basic detector response}. 

The 4th harmonic amplitude terms ($A_4$ and $A'_4$) represent instrumentally induced polarized power. 
This category includes:
\begin{itemize}
\item the stationary signals represented by $A_4$ such as instrument polarized emissions and instrument unpolarized emissions polarized through \ac{IP}. 
%Here \ac{IP} is used in the traditional sense and refers to the conversion of intensity to polarization solely through differential transmission or reflection.
Instrument polarized and unpolarized emission are stationary in that they vary only with thermal variations in the instrument.
As such the $A_4$ term, though it represents the largest polarization signal measured by the detectors (see Section~\ref{section: hwp}), is separable from the sky polarization because it is constant over timescales on which the instrument is thermally stable.
\item scan modulated signals represented by $A'_4 I^{sky}$ which we call in this paper \ac{ICP}. 
\ac{ICP} includes two effects: \ac{IP} acting on $I^{sky}$ that we label \ac{ICP}$^{IP}$, but also \ac{ICP} arising from non-linear detector response which is the subject of this paper and that we label \ac{ICP}$^{NL}$. 
\end{itemize}
 
%However we will show that $A_4$ couples to detector non-linearity to produce \ac{ICP}$^{NL}$, such that minimizing $A_4$ will be important for future experiments.

\noindent In the next section we show the \ac{ICP} observed in EBEX maps. In Section~\ref{section: origin of leakage} we describe in more details the physical mechanisms generating \ac{ICP}$^{IP}$ and \ac{ICP}$^{NL}$.

%The detectors are only sensitive to power and their time-stream $D_t$ is composed of the unpolarized sky signal, the polarized sky signal modulated by the achromatic \ac{HWP}, noise, and a spurious modulation signal $A(\gamma_t)$ coming from \ac{IP} and other spurious Stokes vectors:

%\small
%\begin{align}
%D_t &=& \frac{1}{2} \, \Big (I^{sky}_t + \epsilon Q^{sky}_t \cos(4\gamma_t - \Phi_t) \nonumber \\
%     &+& \epsilon U^{sky}_t \sin(4\gamma_t - \Phi_t) \Big ) \nonumber\\ 
%     &+& A(\gamma_t) + n_t \label{eq: nominal detector model}
%\end{align}
%\normalsize

%\noindent where  

%% file: tex/excess_polarization_maps.tex
\section{Intensity-Coupled-Polarization (ICP) Observed in EBEX Maps}
\label{section: map making}

\input{./tex/instrument.tex}

\subsection{EBEX Maps}
\begin{figure}[h!]
\begin{center}
\begin{tabular}{lr}
\includegraphics[width=0.48\columnwidth]{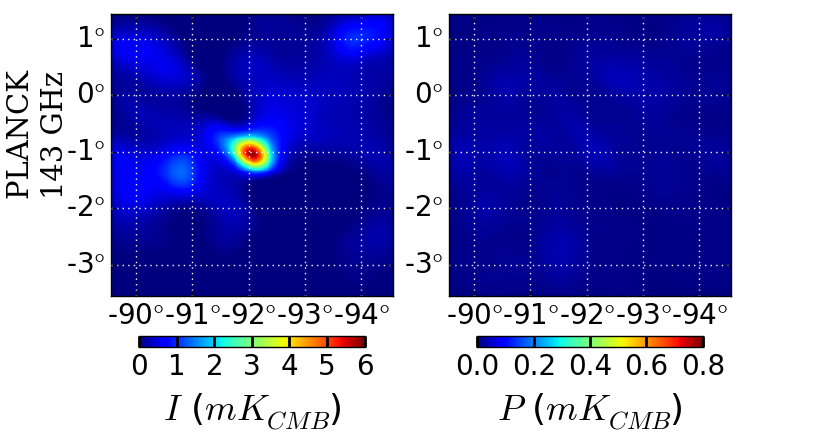} %150
&
\includegraphics[width=0.48\columnwidth]{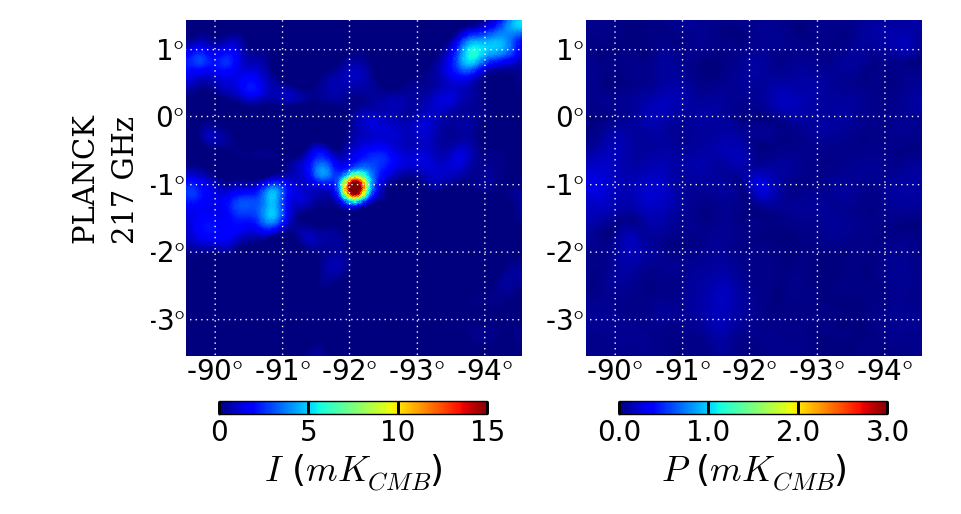} %250
\\
\includegraphics[width=0.48\columnwidth]{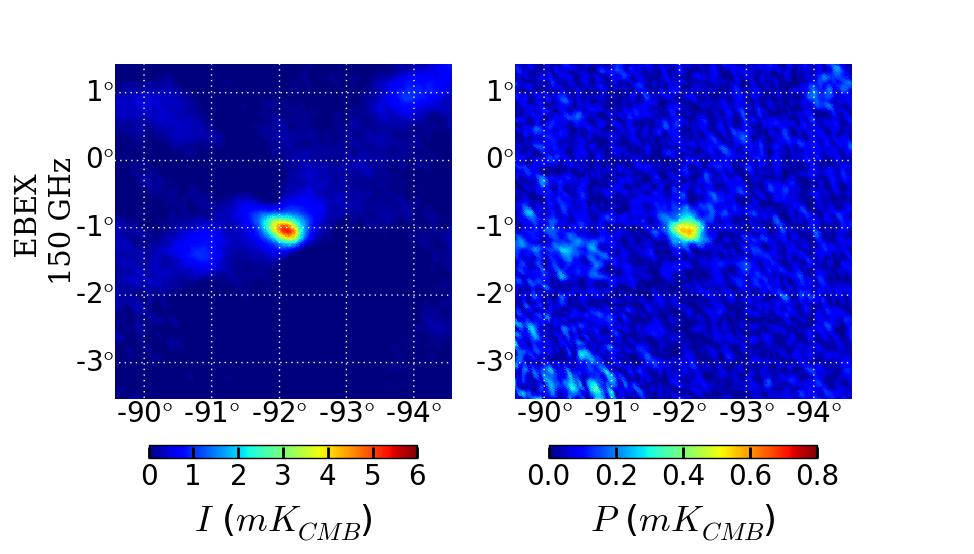} %150
&
\includegraphics[width=0.48\columnwidth]{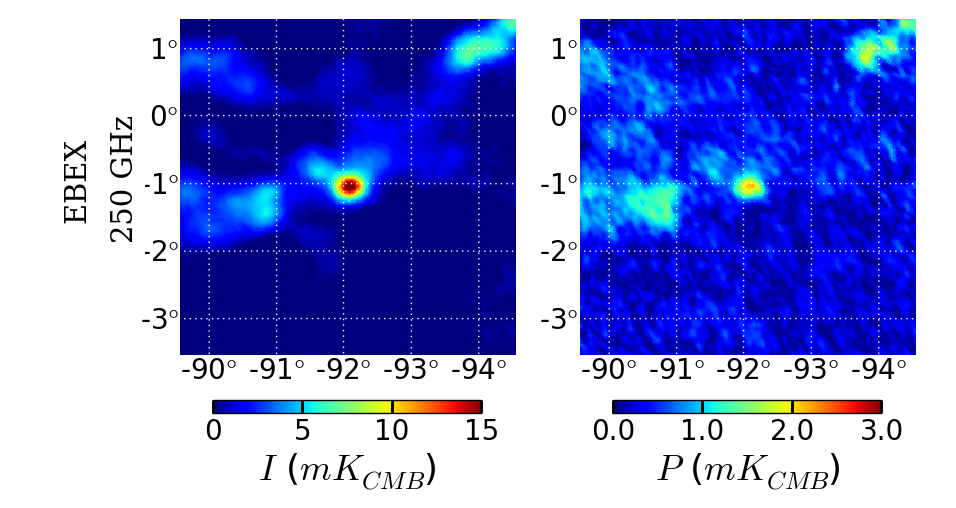} %250

\end{tabular}
\end{center}
\caption{ \footnotesize
Comparison of RCW38 maps in $I$ and $P$ between Planck at 143 GHz (top left), EBEX at 150 GHz (bottom left), Planck at 217 GHz (top right) and EBEX at 250 GHz (bottom right). 
The maps shown in Galactic coordinates co-add 332 (216) detectors at 150 (250) GHz. 
The polarization orientation is reconstructed in the instrument frame. 
%The maps are smoothed with a Gaussian beam of \ac{FWHM} 5~\si{\arcminute}.
}
\label{figure: rcw38}
\end{figure}

%In this section we present sky maps from \ac{EBEX2013} data and show that we observe excess polarization.
We present here maps from EBEX2013 data and show that we observe \ac{ICP}. 
To make maps we remove the stationary part of the \ac{HWP}SS, calibrate, deconvolve the detector time-constant, demodulate and filter the time-streams to extract $I$, $Q$ and $U$ and bin them into pixels. A detailed review of the time-stream processing is available in \cite{joy_thesis} and \cite{derek_thesis}, and the calibration is described in \cite{Aubin_MGrossman2015}. 
%To remove low frequency drifts and high frequency noise, a bandpass filter is applied to $I$, $Q$ and $U$ time-streams with bandwidth \SI{3.0e-2} to \SI{2}{\hertz} for $I$ and \SI{5.0e-2} to \SI{1.5}{\hertz} for $Q$ \& $U$ (after demodulation). 
%Figures~\ref{figure: rcw38 150} and \ref{figure: rcw38 250}  show maps of the bright embedded cluster RCW38 at 150 and 250 GHz for Stokes $I$ and the polarization power $P$ defined as $P = \sqrt{Q^2+U^2}$. 
Figure~\ref{figure: rcw38} shows Planck and EBEX maps of the bright embedded cluster RCW38 for Stokes $I$ and the polarization power $P$, defined as $P = \sqrt{Q^2+U^2}$. 
Polarization orientation is reconstructed in the instrument frame (see definition in Appendix~\ref{appendix: coordinates}). 
%We compare RCW38 between Planck (top) and EBEX2013 (bottom) maps. 
%To generate figures from Planck data, 
Planck maps closest in frequency to the EBEX bands are first smoothed to the EBEX beam size. 
Planck time-streams are then generated using the EBEX pointing and HWP angles, and those time-streams are processed and binned into maps using the same pipeline as EBEX2013 time-streams. 
%An excess polarization power is visible in RCW38 maps at both frequency bands when comparing EBEX2013 to Planck. 
Excess polarization in the EBEX data is apparent for both 150 and 250~GHz maps. 
The $I$ to $P$ Pearson correlation coefficient and linear slope are given in Table~\ref{table: I to P coeffs}. 
The high correlation coefficient between $I$ and $P$ points to the excess polarization in EBEX2013 coming from \ac{ICP}. 
The measured linear slopes of 11\% (12\%) for 150 (250) GHz correspond to a measurement of $A'_4$ (from Equation~\ref{eq: hwp model}) averaged over detectors. 
These numbers are larger than the maximum anticipated \ac{IP} of 2.7\% as will be explained in Section~\ref{subsection: icp_ip}.
%-- as we will show this is because the \ac{ICP} comes from detector non-linearity rather than \ac{IP}.
%The origin of this discrepancy is discussed the following sections. %Note that polarization orientation from an $I$ to $P$ leakage is independent of telescope roll angle $\psi_t$, which is why we plot the orientation in the instrument frame where it adds up coherently.

\begin{table}[h!]
%\scriptsize
\begin{center}       
\makebox[0.8\columnwidth][c]{
\begin{tabular}{l|ll|ll}
               & \multicolumn{2}{c}{RCW 38} & \multicolumn{2}{c}{CMB Stacked Spots}  \\
               \hline
               & Correlation  & Linear     & Correlation  & Linear  \\
               & Coefficient  & Slope (\%) & Coefficient  & Slope (\%)        \\
               \hline
Planck 143 GHz &    0.1      & 0.2     &  0.3  &  0.0 \\
EBEX 150 GHz   &    0.8      & 11      &  0.8  &  8  \\
Planck 217 GHz &    0.3      & 0.7     &  0.2  &  0.1 \\
EBEX 250 GHz   &    0.8      & 12      &  0.6  &  16 \\
        \hline
\end{tabular}
}
\end{center}
\caption{\footnotesize Pearson correlation and linear slope (corresponding to an average of $A'_4$ across detectors) between $I$ and $P$ using RCW38 and stacked CMB maps. 
For RCW38, only pixels with $I$ greater than 2 (9)~\si{\milli\kelvin} are used for 150 (250) GHz calculations. 
For CMB, only pixels with $I$ greater than 10~\si{\micro\kelvin} are used for calculations.
We estimate the one sigma error on the slope to be 1\% for EBEX data and 0.1\% for Planck data.}
\label{table: I to P coeffs}
%\normalsize
\end{table}

%In the 150 GHz maps, selecting for pixels with $I$ greater than 2~\si{\milli\kelvin}, EBEX2013 P and I show a correlation coefficient of 0.74 and a linear slope of 0.133. By comparison, Planck $P$ and $I$ have a correlation coefficient of 0.55 and a linear slope of 0.003. 

%%%%%%%%%% RCW38 maps -- ONE column format, all frequencies in one plot
We ascertain the existence of \ac{ICP} by co-adding maps around \ac{CMB} cold and hot spots \citep{komatsu11}. 
We identify spot locations by examining the Planck CMB maps \footnote{\tiny \url{http://irsa.ipac.caltech.edu/data/Planck/release\_2/all-sky-maps/cmbpreviews/COM\_CMB\_IQU-commander\_1024\_R2.02\_full/index.html}} (see \cite{joy_thesis}). 
We smooth the EBEX $I$, $Q$, $U$ maps to 0.5\si{\degree} ($Q$, $U$ are oriented in the instrument frame) and extract a square region of 5\si{\degree}~$\times$~5\si{\degree} around the spot extremum.  
The hot and cold spots are stacked.
%Starting from sky $I$, $Q$ and $U$ maps smoothed to 0.5\si{\degree} (with $Q$ and $U$ oriented in the instrument frame), we extract a square region of 5\si{\degree}~$\times$~5\si{\degree} around each \ac{CMB} cold or hot spot and co-add the extracted images for $I$, $Q$ and $U$ respectively. The cold and hot spots are identified using the Planck \ac{CMB} 
%maps\footnote{\tiny \url{http://irsa.ipac.caltech.edu/data/Planck/release\_2/all-sky-maps/cmbpreviews/COM\_CMB\_IQU-commander\_1024\_R2.02\_full/index.html}} 
%by looking for local minimas and maximas. Details of the processing can be found in \cite{joy_thesis}. 
Figures~\ref{figure: cmb stacked 150}~and~\ref{figure: cmb stacked 250} show the resulting stacked spots from co-adding $\sim$2000 spots using 150 and 250 GHz detectors, respectively. 
We show the stacked spots in $I$ and also in polarization power $P$ made from the $Q$ and $U$ stacked spots. In both EBEX and Planck, the CMB is visible in the co-added $I$ maps. For polarization co-added in the instrument frame coupled to the EBEX scan strategy, we expect no \ac{CMB} polarization power in the stacked spots, and none is observed in the Planck $P$ data. In EBEX, polarization power is visible in the center of the stacked $P$ map, this is the result of \ac{ICP}. The correlation coefficient and linear slope between $I$ and $P$ are shown in Table~\ref{table: I to P coeffs} and the EBEX numbers are consistent 
between the RCW38 and CMB stacked map measurements. 
%In the next section, we use the Galaxy to measure $A'_4$ and $\alpha'_4$ for each detector .
In the next section, we describe in more details the two mechanisms (IP and detector non-linearity) responsible for \ac{ICP}. 
%and in Section~\ref{section: leakage params characterization} we use EBEX data to show \ac{ICP}$^{NL}$ is the main contributor.

%%%%%%%%%% Stacked CMB maps -- one column format %%%%%%%%%%%%%%%%%%%%%%%%%%%

{\addtolength{\topmargin}{-0.9in}
\begin{figure}[h!]
\begin{center}
\begin{tabular}{c}

\makebox[\columnwidth][c]{
\includegraphics[width=\columnwidth]{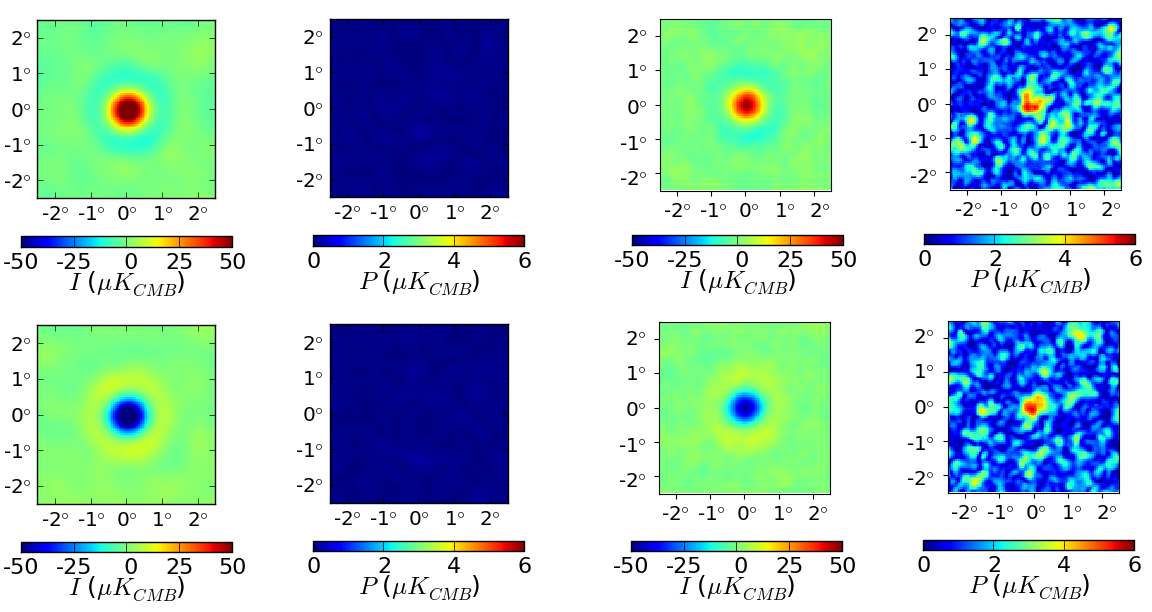}
}
\end{tabular}
\end{center}
\vspace{-25pt}
\caption{\footnotesize Co-added CMB hot and cold spots for Planck 143 GHz (left four) and EBEX 150 GHz (right four) in $I$ and $P$. 
For each experiment, hot (cold) spots are shown on top (bottom) and co-add 2122 (2255) spots.}
\label{figure: cmb stacked 150}
%\end{figure}
%\vspace*{\floatsep}
%%%%%%%%%% Stacked CMB maps 250
%\begin{figure}[h!]
\vspace{-25pt}
\begin{center}
\begin{tabular}{c}
\makebox[\columnwidth][c]{
\includegraphics[width=\columnwidth]{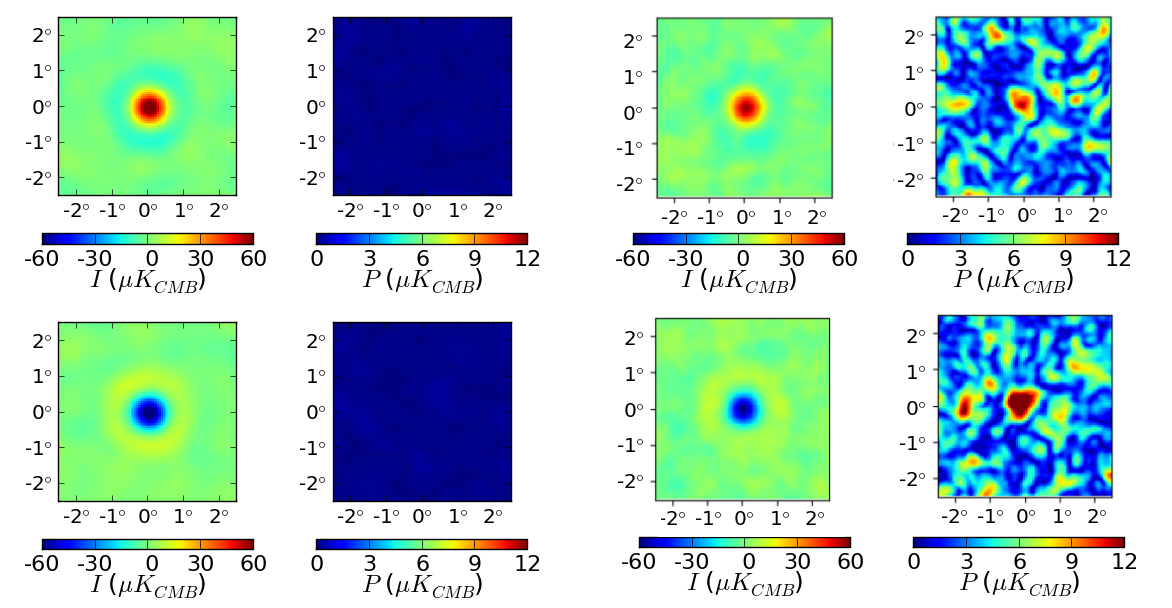}
}
\end{tabular}
\end{center}
\vspace{-25pt}
\caption{\footnotesize Co-added CMB hot and cold spots for Planck 217 GHz (left four) and EBEX 250 GHz (right four) in $I$ and $P$. 
For each experiment, hot (cold) spots are shown on top (bottom) and co-add 1918 (2092) spots.}
\label{figure: cmb stacked 250}
\end{figure}
}

\clearpage

%% file: tex/instrument.tex
%\section{The EBEX Instrument}
\subsection{The EBEX Instrument}

A detailed description of the instrument is available in \cite{EBEXPaper2,EBEXPaper3,EBEXPaper1}.
We provide here a summary relevant to the understanding of the origin of the \ac{ICP}.
%We provide here a summary helpful to understand the origin of the \ac{ICP} discussed
%in this paper.
The \ac{EBEX} instrument was a balloon-borne telescope designed to measure the E- and 
B-mode polarization of the CMB while simultaneously measuring Galactic dust 
emission over the range 30~$<$~$\ell$~$<$~1500 of the angular power spectrum.  
To achieve sensitivity to both the CMB polarization signal and galactic 
foregrounds, EBEX had three bands centered on 150, 250, and 410~GHz. 

\begin{figure*}[h!]
\begin{center}
\begin{tabular}{c}

\includegraphics[width=1.0\textwidth]{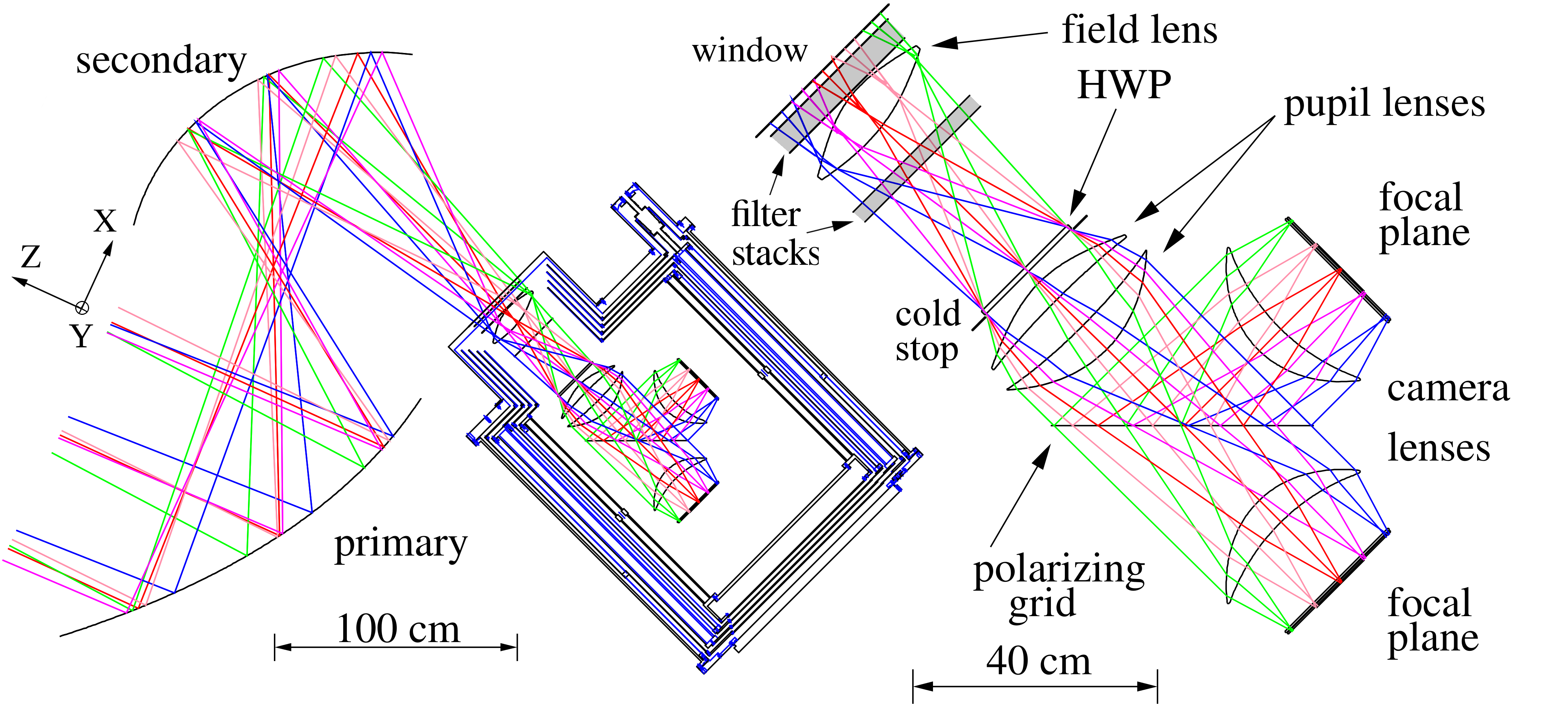}

\end{tabular}
\end{center}
\caption{\footnotesize Ray tracing of the EBEX optical design consisting of two ambient temperature reflectors in a Gregorian configuration and a cryogenic receiver (left). Inside the receiver (right), cryogenically cooled polyethylene lenses formed a cold stop and provided diffraction limited performance over a flat, telecentric, \SI{6.6}{\degree} field of view.}
\label{fig: ray_tracing}
\end{figure*}

The telescope optics comprised warm primary and secondary mirrors and a series of 
cold lenses and filters located inside a cryogenically cooled receiver 
(see Figure~\ref{fig: ray_tracing}).  
%The warm mirrors were situated in an off-axis Gregorian Mizuguchi-Dragone configuration to minimize polarized systematics 
%\citep{hanany_optics}.  Incoming light was reflected from the 1.5~m parabolic 
%primary mirror to the 1.10~m by 0.98~m elliptical secondary mirror, which then 
%reflected the light into the receiver.
%%
%Light entering the receiver cryostat first encountered an 0.51~mm thick, 30~cm diameter 
%vacuum window made of ultra-high molecular weight polyethylene (UHMWPE).  
%The cryostat included five progressively cooler stages: ambient temperature ($\sim$~300~K) 
%at the window; 77~K and 4~K stages cooled by liquid cryogens; and a 1~K inner stage and 
%the 0.25~K detector focal planes cooled by closed-cycle helium adsorption fridges.  
%Low-pass filters at the first three stages reduced thermal loading on the inner stages. 
%%
\ac{EBEX} achieved polarimetry via a stationary wire-grid polarizer and a 24~cm diameter continuously rotating
achromatic \ac{HWP} composed of a stack of five birefringent sapphire disks following a
Pancharatnam design \citep{pancharatnam}.
%The polarizing grid mounted at 45$^\circ$ to the optical path split the beam into independent 
%polarizations and directed them to independent (``horizontal'' and ``vertical'') focal 
%planes.  
Incoming optical rays were focused onto each focal plane by a field lens and a series 
of pupil and camera lenses. The \ac{HWP} was kept at 4~\si{\kelvin} and located
at an aperture stop such that each detector beam covered the \ac{HWP}. The field lens
was located at an image of the focal plane.
Each focal plane contained an array of transition-edge sensor (TES) bolometric detectors 
arranged into seven hexagonal wafers, with four 150~GHz wafers, two~250 GHz wafers, and 
one 410~GHz wafer per focal plane.  
%Each wafer was layered with a band-defining low-pass 
%metal mesh filter.  Conical feed horns and cylindrical wave guides then coupled the incident
%radiation to the detectors. 
EBEX operated 955 detectors during its science flight.

EBEX launched from McMurdo station, Antarctica on December~29, 2012, circumnavigating 
the continent at an altitude of $\sim$~35~km and landing 25~days later on January~23, 2013.  
We refer to data from this flight as EBEX2013.
The cryogenic system that cooled the receiver was active for 11~days before cryogens 
depleted. Due to an error in thermal modelling \citep{EBEXPaper3}, EBEX was unable to point in azimuth
and as a result EBEX scanned a 5,700~deg$^2$ strip of sky delimited by 
declination $-67.9^\circ$ and $-38.9^\circ$, corresponding to free rotations in azimuth
at a constant elevation of 54\si{\degree}.

%During that time it was intended that EBEX would perform science observations 
%on a 400~deg$^2$ patch that was selected for its low foreground contribution.  Due to a 
%thermal modeling error, the pivot motor controller that was responsible for controlling 
%azimuth motion was provided with inadequate radiative shielding.  As a result the pivot motor 
%controller overheated and automatically shut down, allowing the instrument to float freely 
%in azimuth for a majority of the flight.  The telescope's azimuth motion was determined by 
%the rotation of the balloon and the spring constant of the flight train.  The resulting 
%azimuth motion was a superposition of 360$^\circ$ rotations every 15 to 60 minutes and 
%oscillations of variable amplitude with a period of $\sim$~80~s.  
%After loss of azimuthal control the telescope elevation was fixed 
%at 54$^\circ$ to maintain an angular separation of $\sim$~15$^\circ$ between the telescope 
%boresight and the Sun's maximum elevation.  The resulting sky coverage consisted of a 
%5,700~deg$^2$ strip of sky delimited by declination $-67.9^\circ$ and $-38.9^\circ$.  

%The EBEX instrument is described in additional detail in 
%\cite{EBEXPaper2, EBEXPaper3, EBEXPaper1}.

%% file: tex/icp_models.tex
%%%%%%%%%%%%%%%%%%%%%%%%%%%%%%%%%%%%%%%%%%%%%%%%%%%%%%%%%%%%%%%%%
\section{Mechanisms for Intensity-Coupled-Polarization (ICP)}
\label{section: origin of leakage}

%In this section we outline two mechanism, optical \ac{IP} and detector non-linearity, that cause $I \rightarrow P$ leakage and show how we can use the measured leakage parameters to differentiate between them.
We examine here two physical mechanisms responsible for \ac{ICP} and trace the amplitude and polarization angle of each as a function of focal plane position. 
This provides a way to distinguish between them.
%and provide insights as to how the amplitude and polarization angle of each \ac{ICP} varies with focal plane position. 
%This will be used to determine the dominant source of \ac{ICP} in \ac{EBEX}.

\subsection{Instrumental Polarization (IP)}
\label{subsection: icp_ip}

Mirror and lenses sky-side of the \ac{HWP} are typical sources of \ac{IP} in \ac{CMB} instruments. 
Unpolarized radiation $I^{sky}$ incident on the instrument will produce an \ac{ICP}$^{IP}$ signal $I^{sky} \,  \varepsilon^{IP} \, \cos(4\gamma_t - 2 \alpha^{IP})$ that has polarization fraction $\varepsilon^{IP}$ and polarization angle $\alpha^{IP}$, where $\varepsilon^{IP}$ and $\alpha^{IP}$ are determined by the instrument configuration.

In EBEX, the optical design software Code V\footnote{\url{https://optics.synopsys.com/codev/}} shows that the main source of \ac{IP} is the field lens, dominating the mirror \ac{IP} by an order of magnitude at 150 and 250~GHz.
Figure~\ref{fig: ray_tracing} shows the location of and incident rays on the field lens.
The amount of \ac{IP} from the field lens increases with distance $d$ away from the lens center because of the increasing incident angles from the lens curvature.
Unpolarized radiation $I^{sky}$ incident on the lens will be polarized in the plane of incident light (see Appendix~\ref{appendix: dielectric} for a general derivation).
Over all rays hitting the field lens at a given location forming an angle $\beta$ with the x-axis, the outgoing polarization will have a polarization angle $\alpha^{IP} = \beta$.

The \ac{EBEX} field lens is located at an image of the focal plane such that the IP properties directly translate to the focal plane.
%which gives unique properties to the polarization coming from field lens \ac{IP}. 
Let the polar coordinates of a detector on the focal plane be its radial distance from the center $r_{det}$ and its polar angle $\rho_{det}$.
%Because the field lens is at an image of the focal plane, 
The detector illuminated by a ray hitting the field lens at radius $d$ and angle $\beta$ is the detector with coordinates $r_{det} = d$ and $\rho_{det} = \beta$. 
Therefore the \ac{ICP}$^{IP}$ of each detector has polarization angle $\alpha^{IP}$ equal to the polar angle $\rho_{det}$ of the detector position on the focal plane, and polarization fraction $\varepsilon^{IP}$ which increases for a detector at the edge of the focal plane. Code V modelling for EBEX shows a maximum polarization fraction $\varepsilon^{IP}$ of $2.7$~\% at the edge of the focal plane for the 150 and 250~GHz frequency bands \citep{EBEXPaper1}.  

If \ac{IP} is the dominant source of \ac{ICP} in EBEX, we expect $A'_4$ (from Equation~\ref{eq: hwp model}) to be of order $\varepsilon^{IP} \sim 2.7\%$ given Code V predictions , and the \ac{ICP} polarization angle $\alpha'_4$ to be equal to the IP polarization angle $\alpha^{IP}$ which in EBEX is equal to the detector polar angle $\rho_{det}$.
Future experiments wishing to mitigate \ac{ICP}$^{IP}$ can diminish the magnitude of $\varepsilon^{IP}$ by placing the HWP at the beginning of the optical chain: only optical elements sky-side of the HWP contribute to the total IP.

\subsection{Detector Non-Linearity}
\label{subsection: detector nl icp}

Another possible source of \ac{ICP} is detector non-linearity in the presence of a HWPSS with a 4th harmonic.
%$A_4$, as we show in the calculation below.
%Another possible source of \ac{ICP} is detector non-linearity. 
%In EBEX the non-linearity is caused by a strong rotation synchronous signal, parameterized by A_4. This origin gives the ICP unique signatures, which we now derive. 

We derive the properties of the \ac{ICP}$^{NL}$ using a simplified version of the data model in Equation~\ref{eq: nominal detector model} in which the incoming power on the detectors is composed solely of an unpolarized sky signal and a stationary 4th harmonic HWPSS parametrized by $A_4$:

\begin{equation}
D_t = I^{sky}_t + A_4 \cos(4 \gamma_t - 2 \alpha_4) \label{eq: simple model for nl}.
\end{equation}

Let $D_t$ vary over a range larger than the linear range of the detector response. % the response of the detector becomes non-linear.
For this derivation, we limit our non-linearity model to second order terms and ignore time-constant effects. We write the non-linear detector response as

\begin{equation}
D^{NL}_t = f^{NL}(D_t) = D_t - K D_t^2
\end{equation}

\noindent where  $K$ has unit of inverse power and characterizes the non-linearity of the detector. For TES detectors tuned in the high-resistance regime of their superconducting transition, we can assume $K > 0$, as we show in Appendix~\ref{appendix: detector non-linearity}.
We now re-write the detector time-stream as:

\begin{align}
D^{NL}_t = & \, f^{NL}\big(I^{sky}_t + A_4 \cos(4 \gamma_t - 2 \alpha_4)\big) \nonumber \\
         = & \, (1-K I^{sky}_t) I^{sky}_t \nonumber \\
         &+ 2 A_4 K \, I^{sky}_t \, \cos(4\gamma - 2(\alpha_4 + \frac{\pi}{2})) \nonumber \\
         &+ A_4 \cos(4\gamma - 2 \alpha_4) \nonumber \\
         &- \frac{1}{2} K A_4^2 \cos(8\gamma - 4 \alpha_4) - \frac{1}{2} K A_4^2 \label{eq: nl response}.
\end{align}

\noindent The non-linear response has multiple effects:
\begin{itemize}
\item it decreases $I^{sky}_t$ by $(1-K I^{sky}_t)$;
\item it creates an \ac{ICP}$^{NL}$ signal $2 A_4 K I^{sky}_t \cos(4\gamma - 2(\alpha_4 + \frac{\pi}{2}))$, with polarization fraction $\varepsilon^{NL} = 2 A_4 K$ and 
polarization angle $\alpha^{NL} = \alpha_4 + \frac{\pi}{2}$;
\item it creates higher harmonics in the HWPSS (in this second order example, only an 8th harmonic), as well as modify the DC level.
\end{itemize}

\noindent Our model does not include intrinsic sky polarization $P^{sky}_t$, but one can show similarly that non-linearity decreases $P^{sky}_t$ by $(1-2 K I^{sky}_t)$. 

If non-linearity is the dominant source of \ac{ICP} in EBEX, we expect $A'_4$ to be of order $\varepsilon^{NL} = 2 A_4 K$, and the \ac{ICP} polarization angle $\alpha'_4$ to be equal to the non-linear model polarization angle $\alpha^{NL}$ which is offset from the stationary HWPSS 4th harmonic polarization angle $\alpha_4$ by $\frac{\pi}{2}$.
Note that because the polarization fraction of \ac{ICP}$^{NL}$ is determined by the product of $A_4$ and the detector non-linearity $K$, future experiments wishing to minimize \ac{ICP}$^{NL}$ can act on both the non-linearity of the detectors ($K$) and the magnitude of the stationary HWPSS 4th harmonic ($A_4$), the latter by minimizing the polarized and unpolarized thermal emissions sky side of the HWP.

To determine the origin of the ICP observed in EBEX, we can measure the ICP polarization angle $\alpha'_4$ and compare it to both $\rho_{det}$ and to the stationary HWPSS polarization angle $\alpha_4$. In Section~\ref{section: leakage params characterization} we use the data to show that the ICP polarization angle $\alpha'_4$  is consistent with a non-linear origin of the signal, and is not consistent with IP as its origin.   
We first determine the properties of the stationary HWPSS.

%% file: tex/hwp_template.tex
\section{Stationary HWP Synchronous Signal}
\label{section: hwp}

\begin{figure}[h!]
\begin{center}
\begin{tabular}{c}
\makebox[\columnwidth][c]{
\includegraphics[width=\columnwidth]{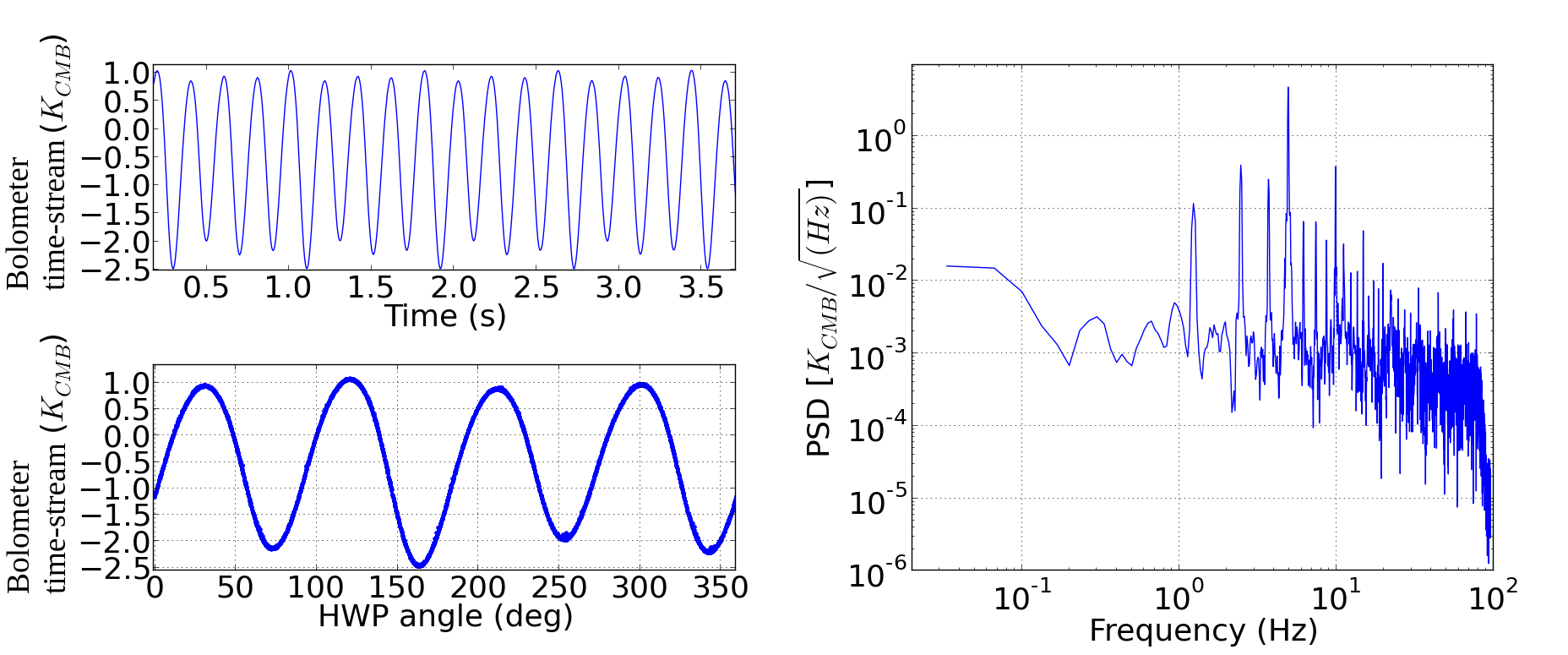}
}
\end{tabular}
\end{center}
%\caption{{\it Top}: Plot of a calibrated detector time-stream over 3.5~\si{\second} of data prior to \ac{HWP}SS removal. The \ac{HWP}SS has amplitude of 3.5~\si{\kelvin} and dominates the signal. {\it Middle}: Plot of the same detector time-stream versus \ac{HWP} angle, showing the \ac{HWP}SS is synchronous with the \ac{HWP} rotation. The 4th harmonic dominates the HWPSS. {\it Bottom}: PSD of the same detector time-stream.}
\caption{\footnotesize {\it Left, top}: Plot of a calibrated detector time-stream over 3.5~\si{\second} of data prior to stationary \ac{HWP}SS removal. The \ac{HWP}SS has amplitude of 3.5~\si{\kelvin} and dominates the signal. {\it Left, bottom}: Plot of the same detector time-stream versus \ac{HWP} angle, showing the \ac{HWP}SS is synchronous with the \ac{HWP} rotation. The 4th harmonic dominates the HWPSS. {\it Right}: PSD of the same detector time-stream.}
\label{figure: hwp vs time and frequency}
\end{figure}

%In this section we describe the EBEX HWPSS and propose a model for its origin. 
%This is useful to understand the nature of the excess polarization observed in maps, as we will show in further sections. 
%We discuss here sources of the stationary HWPSS 4th harmonic (described by $A_4$ and $\phi_4$) and use EBEX data to asses the validity of our models. 
%Being able to predict $A_4$ is important to design experiments with negligible \ac{ICP}$^{NL}$.
In Figure~\ref{figure: hwp vs time and frequency} we plot a detector time-stream and \ac{PSD} from EBEX, showing that a \ac{HWP}SS dominates the detector time-streams. The HWPSS has power at all harmonics of the \ac{HWP} rotation up to the Nyquist frequency, with the 4th harmonic being the dominant harmonic by an order of magnitude. The stationary part of the \ac{HWP}SS (coefficients $A_j$, $\alpha_j$ in Equation~\ref{eq: hwp model}) is fitted using a maximum likelihood estimator. We refer the reader to \cite{joy_thesis} and \cite{derek_thesis} for a detailed review on the stationary HWPSS fitting and removal. The power in $A_4$ comes from two sources sky side of the \ac{HWP}: unpolarized power (thermal instrument emission, \ac{CMB} monopole and atmosphere) getting polarized through \ac{IP}, as well as polarized thermal emission from the instrument.
%In the next paragraphs we show the EBEX HWPSS is dominated by the former.

%%%%%%%%% bolo.signals vs time, angle and freq

\subsection{Unpolarized Thermal Emission Polarized Through IP}
\begin{figure}[h!]
\begin{center}
\begin{tabular}{c}

\makebox[\columnwidth][c]{
\includegraphics[width=0.7\columnwidth]{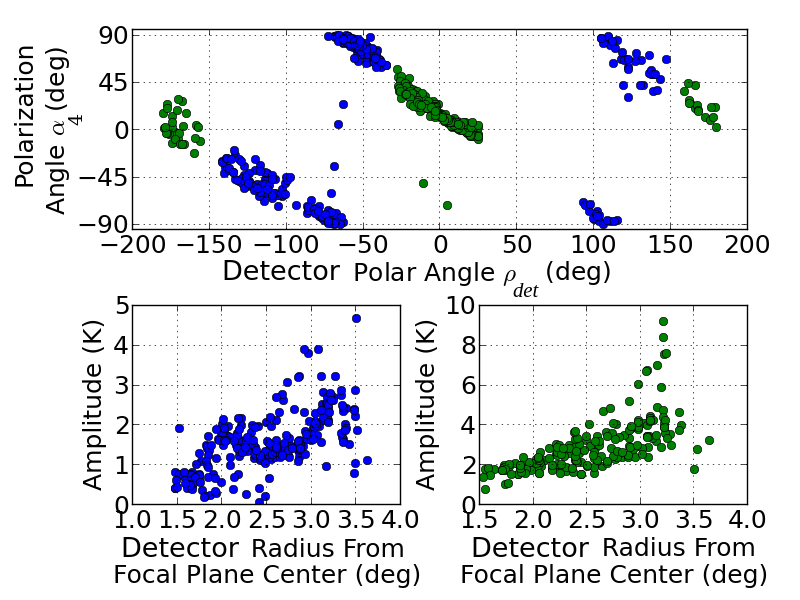}
}
\end{tabular}
\end{center}
\caption{\footnotesize {\it Top:} Polarization angle $\alpha_4$ of the stationary \ac{HWP}SS 4th harmonic for 150 (blue) and 250 (green) GHz detectors, plotted against the detector polar angle $\rho_{det}$ on the focal plane, showing strong 1:1 linear correlation between $\alpha_4$ and $\rho_{det}$ (the negative slope comes from using different conventions for $\alpha$ and $\rho_{det}$). {\it Bottom:} Amplitude $A_4$ of the stationary \ac{HWP} 4th harmonic for 150 (left) and 250 (right) GHz detectors, plotted against the detector distance $r_{det}$ from the focal plane center. Within each frequency band, the amplitude of the HWPSS increases with detector radius.}
\label{figure: HWP 4th vs fp}
\end{figure}

In Section~\ref{subsection: icp_ip} we showed how \ac{IP} acts on $I^{sky}$ to produce \ac{ICP}$^{IP}$. Similarly, \ac{IP} will act on $I^{instr}$ to produce a stationary polarized signal. As discussed earlier, in EBEX because the dominant source of \ac{IP} is the field lens located at an image of the focal plane, polarization signals generated by \ac{IP} will exhibit a distinctive pattern as a function of focal plane position: the polarization angle will be equal to the polar angle of the detector, and the polarized power will increase with radial distance away from the focal plane center ($A_4$ from \ac{IP} is equal to $I^{instr} \varepsilon^{IP}$) .
This is what we observe in the stationary HWPSS 4th harmonic, as shown in Figure~\ref{figure: HWP 4th vs fp} (top and bottom panel). This data indicates field lens \ac{IP} is a dominant source of the stationary HWPSS 4th harmonic. 
Its magnitude is estimated in Table~\ref{tab:results} by combining the thermal load on the detectors measured from flight with the IP predicted from Code V. 
We note that the load measured in flight was larger than what was predicted pre-flight.
We hypothesize that the excess load comes from spillover onto warm, highly emissive surfaces around the mirrors, caused by diffraction around the aperture stop. 
The predicted and measured loads and a discussion of this effect are available in \cite{EBEXPaper2}. 
The excess load increased the amount of unpolarized light passing through the field lens and therefore the HWPSS.
%In \cite{EBEXPaper2} we calculate this excess load comparing the predicted radiative loads to the absorbed power determined by load curves in flight. 
%The bottom panels show \ac{HWP}SS magnitude increases towards the edge of the focal plane, which is also expected from the field lens model in which the HWPSS magnitude is $A_4 = I^{instr} \varepsilon^{IP}$. 

%Furthermore, we estimate $I^{instr}$ by using the emissivity of the optical components sky side of the field lens (mirror, windows, filters), as well as the \ac{CMB} monopole, and multiply this estimate with $\varepsilon^{IP}$ predicted from Code~V. The observed 1-4~\si{\kelvin}$_{CMB}$ magnitude of the 4th harmonic is in agreement with those estimates. 
%See \cite{EBEXPaper1} for detailed contributions to the absolute HWPSS magnitude 

%%%%%%%%%% HWP angle vs focal plane position

\subsection{Polarized Thermal Emission}

Polarized thermal emission from the instrument also contributed to $A_4$. The dominant contribution of polarized thermal emission comes from the mirrors.
%, especially the primary mirror. 
%Polarized thermal emission are calculated by multiplying the 300~K mirror black body with the mirror emissivity and the mirror polarization fraction of thermal emission. 
%Following the conclusions of \cite{strozzi}, we see that the polarization fraction of thermal emission, $\varepsilon^{emis}$, 
%depends on the angle of emission, $\theta$, measured relative to the normal, as:
The polarized fraction of mirror thermal emission $p^{emission}$ as a function of the angle of emission $\theta$ with respect to normal incidence is \citep{strozzi}:

\begin{align}
   p^{emission} &= \frac{\sin^2(\theta)}{1+\cos^2(\theta)}
\end{align}

\noindent The range of angles of emission that couple to our detectors is identical to the range of 
incidence angles for our optics. For the primary mirror, this range is from $10^{\circ}$ to
$45^{\circ}$, giving non-negligible polarization fractions.  We average the polarization 
fraction across the beam for each of the mirrors, finding values for $p^{emission}$ of 
%15.6\% 
16\% and 6.4\% at the center of the focal plane for the primary and secondary mirrors, 
respectively. The polarization angle $\alpha^{emis}$ of the polarized emission should be approximatively 
uniform across the focal plane, with $\alpha^{emis} \approx 0^{\circ}$. 
We observe $\alpha^{emis}$ to be non-zero in the top panel of Figure~\ref{figure: HWP 4th vs fp}, indicating that polarized thermal emission is a sub-dominant contribution
to the stationary HWPSS 4th harmonic.

\subsection{Comparing Measurements to Model Predictions}

Our estimate for the two contributions to the HWPSS (\ac{IP} and polarized emission) is given in Table~\ref{tab:results} along with the observed size of the HWPSS, all given in units of power incident on the telescope. We note that the HWPSS varies across detectors and we only provide here average measurements and predictions. In particular, the two contributions will add differently for different locations across the focal plane given the varying polarization angles.
%Our predictions show polarized thermal emission should dominate the HWPSS, which is in contradiction with the observed dependence of polarization angle $\alpha_4$ on polar angle $\rho_{det}$ on the focal plane at 150 and 250~GHz. 
%The observed HWPSS amplitude is larger than what we predict.
%The predicted HWPSS amplitude from IP including this excess load is shown in the fifth column of Table~\ref{tab:results}.

\begin{table}[h]
  \begin{center}
    \begin{tabular}{|c|c|c|c|c|}
      \hline
                     &  Estimated HWPSS    & Predicted HWPSS     &                 \\ 
      Frequency Band &  size from field    & size from polarized & Observed HWPSS  \\  
            (GHz)    &   lens IP using     &    emission (fW)    &  size (fW)      \\
                     &   flight load (fW)  &                     &                 \\ \hline
             150     &       370           &       85            & 570           \\ \hline
             250     &       720           &       190           & 670           \\ \hline
             410     &       350           &       400           & 560           \\ \hline
%             150     &       0.022        &       0.044         & $0.158 \pm 0.044$  \\ \hline
%             250     &       0.021        &       0.048         & $0.074 \pm 0.021$  \\ \hline
%             410     &       0.005        &       0.023         & $0.020 \pm 0.006$  \\ \hline
    \end{tabular}
    \caption{\small HWPSS 4th harmonic amplitudes predictions and observations, expressed as power incident on the telescope. The conversion from power to CMB temperature is 3.24, 
4.54 and 16.1 $mK_{CMB} / fW$ at 150, 250 and 410 GHz. The two sources of HWPSS don't necessarily have the same polarization angle.
} 
    \label{tab:results}
  \end{center}
\end{table}

%Using an index of refraction $n \sim 1.5$ for the lens, and ignoring the AR coating, we calculate that at the edge of the lens where the effect is maximum, $\eta \sim 1.8$ and $\varepsilon \sim 0.2$, i.e. the maximum \ac{IP} is $\frac{1}{2} \varepsilon \sim  10\%$. 

% \noindent where $D^{IP}$ is the detector time-stream resulting from field lens IP acting on unpolarized incoming light $I^{in}$. 

%In conclusion, the way in which the polarization angle and magnitude of the HWPSS 4th harmonic vary with focal plane position is a strong indicator that the field lens \ac{IP} is the dominant source of the HWPSS. Such a dependence is unlikely to occur from any other \ac{IP} sources or polarized emission sources not located in the vicinity of a focal plane image. 
%Finally we note that we have ignored in Equation~\ref{equation: detector ip} the effect of
%incoming polarized emission $Q^{instr}$ and $U^{instr}$ on the field lens. This will create $P \rightarrow I$ leakage and change the DC level of the detector signal
%which is ignored in our analysis.
%and compress the incoming polarization by $\eta$.
%If the dominant source of the HWPSS 4th harmonic were polarized emissions from the mirrors, such dependence on focal plane position would not be observed. 

%% file: tex/characterization.tex
\section{Single Detector Characterization of ICP}
\label{section: leakage params characterization}

In this section we present a general method to characterize \ac{ICP} coupling coefficients $A'_4$ and $\alpha'_4$ for each detector, independently of the \ac{ICP} origin.
The measurement of the coupling coefficients can inform the physical origin of the \ac{ICP} and be used to remove the excess polarization.

In Equation~\ref{eq: nominal detector model} we showed that the power incident on a detector is the sum of the sky signal and the HWPSS, itself composed of a stationary term
and a term modulated by the sky intensity $I_t^{sky}$. Having removed the stationary HWPSS term and now focusing on the dominant 4th harmonic, the detector 
time-stream becomes: 

\begin{align}
 D^{TOTAL}_t  &\sim \frac{1}{2} \, \Big (I^{sky}_t + P^{sky}_t \cos(4\gamma_t - 2 \psi_t - 2 \alpha^{sky}_t) \Big) + A'_4 I^{sky}_t \cos(4 \gamma_t - 2 \alpha'_4) + n_t \\
              &= D^{sky}_t + D^{ICP}_t + n_t \label{equation: model for excess polarization}
\end{align}

\noindent where $D^{ICP}_t = A'_4 I^{sky}_t \cos(4 \gamma_t - 2 \alpha'_4)$ stands for the \ac{ICP} term, $\psi_t$ is the Galactic roll angle (see Appendix~\ref{appendix: coordinates} for the transition from using $\Phi_t$ to $\psi_t$) and $n_t$ is the noise. 
We note that the polarization of $D^{ICP}_t$ originates in the instrument frame in contrast to the polarization of $D^{sky}_t$ which originates in the sky frame (hence its dependence on the Galactic roll angle $\psi_t$).

%The ICP coupling coefficients can be expressed as
%
%\begin{align}
%A'_4 = \frac{\sqrt{Q_{ICP}^2 + U_{ICP}^2}}{I_{sky}} \\
%\alpha'_4 = \frac{1}{2} \atantwo(U_{ICP}, Q_{ICP})
%\end{align}
%
%\noindent where we have used the usual transformation $Q_{ICP} = A'_4 I^{sky} \cos(2 \alpha'_4)$ and $U_{ICP} = A'_4 I^{sky} \sin(2 \alpha'_4)$. 
To isolate and measure $D^{ICP}$, we make {\bf single detector} $I$, $Q$ and $U$ maps of $D^{TOTAL}$ in the instrument frame. 
The value of each pixel $p$ is:
%Each pixel holds to first order in $\varepsilon^{ICP}$ and after demodulation and filtering:

\normalsize
\begin{align}
I_p &= \frac{\sum\limits_{t} w_t I^{sky}_t} {\sum\limits_{t} w_t} + n^I_p = I^{sky}_p + n^I_p  \\
Q_p &= Q^{sky}_p \frac{\sum\limits_{t} w_t \cos(2\psi_t)} {\sum\limits_{t} w_t} + I^{sky}_p A'_4 \cos(2 \alpha'_4) + n^Q_p \label{eq: q per pixel}  \\
% &\sim \varepsilon^{ICP} \cos(2 \alpha^{ICP}) I^{sky} \\ \nonumber\\
U_p &= U^{sky}_p \frac{\sum\limits_{t} w_t \sin(2\psi_t)} {\sum\limits_{t} w_t} + I^{sky}_p A'_4 \sin(2 \alpha'_4) + n^U_p \label{eq: u per pixel}
% &\sim \varepsilon^{ICP} \sin(2 \alpha^{ICP}) I^{sky} 
\end{align}
\normalsize

\noindent where the summation is over all time samples $t$ pertaining to a given pixel $p$, $w_t$ are the map-making weights and $n^{[I, Q, U]}_p$ is the pixel noise. 
Here we used the usual transformation $Q^{sky} = P^{sky} \cos(2 \alpha^{sky})$ and  $U^{sky} = P^{sky} \sin(2 \alpha^{sky})$. 
To measure the coupling parameters, an unpolarized source can be used ($Q^{sky}_p = U^{sky}_p = 0$) or a polarized source can be sampled with varied Galactic roll such that 
$\sum w_t \cos(2\psi_t)$ and $\sum w_t \sin(2\psi_t)$ tend to zero.
%Ideally, one has access to an unpolarized source with sufficient signal to noise to be visible by single detectors. 
The coupling parameters for each detector are estimated from the maps using ensemble averages of $I_p$, $Q_p$ and $U_p$:

\begin{align}
\tilde{A}'_4 = \frac{\sqrt{\langle Q_p \rangle^2 + \langle U_p \rangle^2}}{\langle I_p \rangle} \label{equation: ip parameters} \\
\tilde{\alpha}'_4 = \frac{1}{2} \arctan(\frac{\langle U_p \rangle}{\langle Q_p \rangle}) \nonumber 
\end{align}

\noindent where the tilde denotes the measured quantity. For EBEX, the three possible sources are the CMB, RCW38 and the Galactic plane. 
EBEX doesn't have enough sensitivity to measure the CMB with single detectors. RCW38 is sampled with enough signal to noise but poor coverage. This leaves the Galaxy which has intrinsic polarization. 
%Sky polarization can be averaged out when reconstructing the polarization in the instrument frame if varied Galactic roll sampling nulls the $\sum w_t \cos(2\psi_t)$ term in Equation~\ref{eq: q per pixel} (and equivalent sinus term in Equation~\ref{eq: u per pixel}).
For an extended source like the Galaxy, summing all the pixels within the source will increase the signal to noise and the sampling of $\psi_t$. 
Using simulations, we estimate the error on the coupling parameters coming from partial Galactic roll coverage to be 1.7\% for $\tilde{A}'_4$ and 3\si{\degree} for 
$\tilde{\alpha}'_4$ for the EBEX scan strategy.

\begin{figure}[h!]
\begin{center}
\begin{tabular}{c}
\makebox[\columnwidth][c]{
\includegraphics[width=0.7\columnwidth]{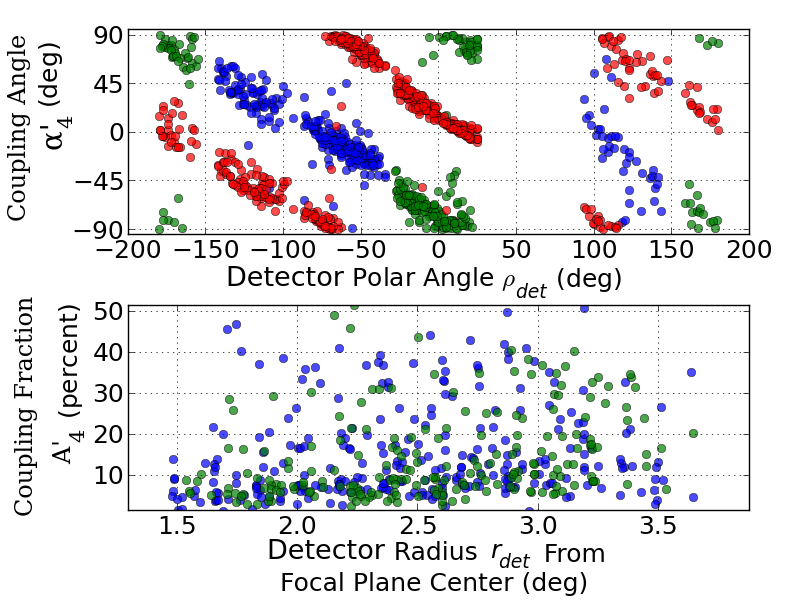}
}
\end{tabular}
\end{center}
\caption{\footnotesize {\it Top:} Measurement of the coupling angle $\tilde{\alpha}'_4$ using the Galaxy for 150 (blue) and 250 (green) GHz detectors, plotted against the detector polar angle $\rho_{det}$. For reference, the HWPSS 4th harmonic polarization angle $\alpha_4$ is also plotted (red). {\it Bottom:} Measurement of the coupling fraction $\tilde{A}'_4$ using the Galaxy for 150 (blue) and 250 (green) GHz detectors, plotted against the detector radius from the focal plane center $r_{det}$.}
\label{figure: ip parameters from galaxy}
\end{figure}

For each detector, we produce $I$, $Q$ and $U$ maps of the Galactic plane in the instrument orientation, with Healpix NSIDE 256 \citep{healpix}. 
We define as valid the pixels located within $\pm$3\si{\degree} of the Galaxy, and with Stokes $I$ value greater than or equal to 3 (15)~\si{\milli\kelvin} for 150 (250)~\si{\giga\hertz} detectors. 
We calculate for each detector the ensemble average $ \langle I \rangle $, $ \langle Q \rangle $ and $\langle U \rangle $ value by averaging all the valid pixels. 
Using those values and Equation~\ref{equation: ip parameters} we estimate for each detector $\tilde{A}'_4$ and $\tilde{\alpha}'_4$, and plot the results in Figure~\ref{figure: ip parameters from galaxy}. 
We observe that the coupling angle $\tilde{\alpha}'_4$ varies linearly with the detector polar angle $\rho_{det}$, and that the coupling fractions $\tilde{A}'_4$ are spread over a wide range with a mode of 7\% and a median absolute deviation of 5.7\%. 
% mean is 17.7\%
These $\tilde{A}'_4$ values are consistent with the RCW38 and CMB linear slopes reported in Table~\ref{table: I to P coeffs} corresponding to an average of $A'_4$ over all detectors. 
%In the following section, we use the coupling parameters to probe the origin of the \ac{ICP}.

%%%%%%%%%%%%%% Leakage Parameters vs FP position + HWP polarization angle
%%%%%%%%%%%%%%%%%%%%%%%%%%%%%%%%%%%%%%%%%%%%%%%%%%%%%%%%%%%%%%%%%%%%%%%%

\subsection{Origin of ICP in EBEX}

%The set of coupling fraction $\varepsilon^{ICP}$ observed in the bottom plot of Figure~\ref{figure: ip parameters from galaxy} shows a large dispersion and values in the range 1-20\% which can be explained by different detectors exhibiting different non-linear responses. 
Comparing the measured polarization angle $\tilde{\alpha}'_4$ to the stationary HWPSS polarization angle $\alpha_4$ is 
a good way to determine the origin of the \ac{ICP}. 
For \ac{ICP}$^{IP}$ the two angles should have the same phase given that both the \ac{ICP} and the stationary HWPSS originate from \ac{IP}. 
In the \ac{ICP}$^{NL}$ case the two angles should be offset by $\frac{\pi}{2}$ as we showed in Section~\ref{subsection: detector nl icp}. 
Furthermore, the coupling fraction $A'_4$ for the \ac{IP} model should be
of order 2.7\% as predicted by Code V, whereas in the non-linear model $A'_4$ is proportional to the HWPSS 4th harmonic amplitude $A_4$
and the amount of non-linearity $K$. Finally, a second-order non-linear response gives rise to an 8th harmonic in the HWPSS, and more complex non-linear response will give rise 
to a multitude of higher harmonics in the HWPSS. The presence of those harmonics can be checked in the HWPSS data. 
We note that higher harmonics can also come from temperature and thickness variations in the HWP and are not necessarily a consequence of non-linearity in the detectors.

Figure~\ref{figure: ip parameters from galaxy} (top panel) shows the measured coupling angle $\tilde{\alpha}'_4$ as a function of the detector polar angle $\rho_{det}$, 
as well as the stationary HWPSS 4th harmonic polarization angle $\alpha_4$. 
The coupling angle $\tilde{\alpha}'_4$ varies linearly with the detector polar angle $\rho_{det}$, which is expected in both the \ac{IP} model and the non-linear model. 
The two sets of angles are offset by $\pi/2$, indicating that the non-linear effect is likely to be the dominant source of \ac{ICP}. 
Additional support for the model that ICP$^{NL}$ is dominant comes from the bottom panel of Figure~\ref{figure: ip parameters from galaxy} showing that the coupling fractions $\tilde{A}'_4$ are spread over a wide range, and on average larger than the maximum $\varepsilon^{IP}$ of 2.7\%  calculated by the Code V simulation.
Finally, a strong 8th harmonic and a multitude of higher harmonics are observed in the EBEX HWPSS as can be seen in Figure~\ref{figure: hwp vs time and frequency}. 
%This also is expected in the non-linear model.

To summarize, the observed properties of the HWPSS and the \ac{ICP} point to the following model. 
Unpolarized instrument emissions are polarized through differential transmission by the field lens and cause a 4th harmonic in the HWPSS with a large amplitude and a polarization angle $\alpha_4$ that varies linearly with the detector polar angle $\rho_{det}$. The magnitude of the HWPSS induces a non-linear response in the detectors which is synchronous with the HWPSS.
The non-linear response couples unpolarized sky signal into the polarization signal bandwidth. 
The polarization angle $\alpha'_4$ of the coupling is offset by $\pi /2 $ from the HWPSS 4th harmonic polarization angle. The non-linear response explains why the observed coupling fractions $\tilde{A}'_4$ are larger than those predicted by optical simulations in Code V and contributes to higher harmonics in the HWPSS. In the next section, we use the measured coupling parameters to remove the spurious polarization in the time domain.

%A note on the validity of our model. 
%When treating both the \ac{IP} case and the non-linear case, we assumed no sky sky polarization. When sky polarization is present, the sky and spurious polarization signals add up coherently in the time domain, though the sky polarization is fixed in the celestial sphere while the coupling polarization is fixed in the instrument frame. 
%The full equations including sky polarization are treated in Appendix~\ref{appendix: non-linearity with sky polarization}. \comred{[Do we need this? All it shows is that to first order our analysis ignoring sky P is valid]}. 
%When including higher order non-linear terms, this will produce higher harmonics in the HWPSS as well as coupling from $I^{sky}$ to higher order terms. In the next section, we show through simulations that modelling the excess polarization as $\frac{1}{2} \varepsilon^{ICP} \cos(4 \gamma_t - 2 \alpha^{ICP})$ (i.e. approximating the non-linear response to a second-order term) is a valid approximation given the measured amount of non-linearity in EBEX.

%% file: tex/icp_removal.tex
%%%%%%%%%%%%%%%%%%%%%%%%%%%%%%%%%%%%%%%%%%%%%%%%%%%%%%%%%%%%%%%%%
\section{Removal of ICP}

%We seek to remove the excess polarization from EBEX maps. 
%In the previous sections we showed that to first order, ICP can be modeled by $D^{ICP}_t \equiv A'_4 I^{sky}_t \cos(4\gamma_t - 2 \alpha'_4)$  regardless of the origin of the coupling (\ac{IP} or detector non-linearity). 
Having measured the coupling parameters, we now produce corrected time-streams $D^C_t$ for each detector:

\begin{equation}
D^C_t = D^{TOTAL}_t - \tilde{D}^{ICP}_t  \label{eq: cleaned timestreams}
\end{equation}

\noindent where $D^{TOTAL}_t$ is the measured detector time-stream including ICP (see Equation~\ref{equation: model for excess polarization}) and $\tilde{D}^{ICP}_t = \tilde{A}'_4 I^{sky}_t \cos(4\gamma_t - 2 \tilde{\alpha}'_4)$ is the measured ICP.

To produce $\tilde{D}^{ICP}_t$, we use for each detector the measured parameters $\tilde{A}'_4$ and $\tilde{\alpha}'_4$ plotted in Figure~\ref{figure: ip parameters from galaxy}. 
Alternatively, in the case of \ac{ICP}$^{NL}$ we can compute $\tilde{\alpha}'_4$ from the stationary HWPSS 4th harmonic polarization angle: $\tilde{\alpha}'_4 = \alpha_4 + \pi/2$. 
The two methods produce similar results. 
The latter method has the advantage that if the HWPSS angle $\alpha_4$ varies over time, $\alpha'_4$ will vary accordingly and this will be reflected by using $\tilde{\alpha}'_4 = \alpha_4 + \pi/2$. $I^{sky}_t$ is generated using the detector pointing and a reference $I$ map (either an EBEX map or  Planck components maps integrated over the EBEX frequency bandwidth). 
Finally, we make EBEX $Q^{sky}$ and $U^{sky}$ maps using the corrected time-streams $D^C_t$ with the same pipeline that was presented in Section~\ref{section: map making}. 
In the following subsections, we present RCW38 maps and CMB stacked spots generated from the cleaned time-streams. 
We present results both in simulations and on EBEX2013 data.

\subsection{Simulations}

We use simulations to evaluate the ICP removal method. Though the method removes ICP from any source, our simulations focus on ICP$^{NL}$ because it is the dominant
source in EBEX. We compare maps made from three datasets:
\begin{enumerate}
\item a ``reference" dataset, obtained from scanning an input sky with detectors that have a linear response (hence no \ac{ICP}$^{NL}$).
\item a ``non-linear" dataset, obtained from scanning the same input sky with detectors that have non-linear response.
\item an ``ICP removed" dataset, obtained from scanning the same input sky with detectors that have non-linear response and then applying the ICP removal technique described earlier. 
\end{enumerate}

\noindent We simulate detector time-streams as follows:

\begin{align}
    D^{SIM}_t  &= f^{NL} \Bigg[ \frac{1}{2} \, \Big (I^{sky}_t + Q^{sky}_t \cos(4\gamma_t - 2 \psi_t) \nonumber +  U^{sky}_t \sin(4\gamma_t - 2 \psi_t) \Big ) \nonumber \\
     & \qquad + \sum \limits_{j=1}^{j=4} A_{j} \cos(j \gamma_t - 2 \alpha_{j}) + n_t  \Bigg] \label{eq: simulated nl}
\end{align}

\noindent $I^{sky}_t$, $Q^{sky}_t$ and $U^{sky}_t$ are generated by scanning input maps with the EBEX scan strategy. 
The input maps come from the Planck Sky Model integrated over the EBEX bandwidth and smoothed to the EBEX beam size. 
We present here the simulations for the 250 GHz detectors.
The non-linear response $f^{NL}$ is set to identity to simulate the reference dataset. 
For the non-linear and ICP removed datasets, we use the following polynomial estimated from EBEX data \citep{derek_thesis}:

\begin{equation}
f_{NL}(D_t) = D_t - 0.04 D_t^2 + 0.001D_t^3
\end{equation}

\noindent The stationary HWPSS is simulated using only the largest physically motivated harmonics 1, 2 and 4, though after the non-linear response is applied multiple higher harmonics are present. 
The HWP coefficients $A_j$ and $\alpha_j$ are sampled from EBEX2013 data ensuring the simulated HWPSS has similar amplitude and focal plane dependence as the EBEX HWPSS. 
EBEX-level white noise $n_t$ is added except when otherwise noted.

\begin{figure}[h!]
\centering     %%% not \center
\makebox[\textwidth][c]{
\subfigure[``Reference"]{\includegraphics[width=0.38\columnwidth]{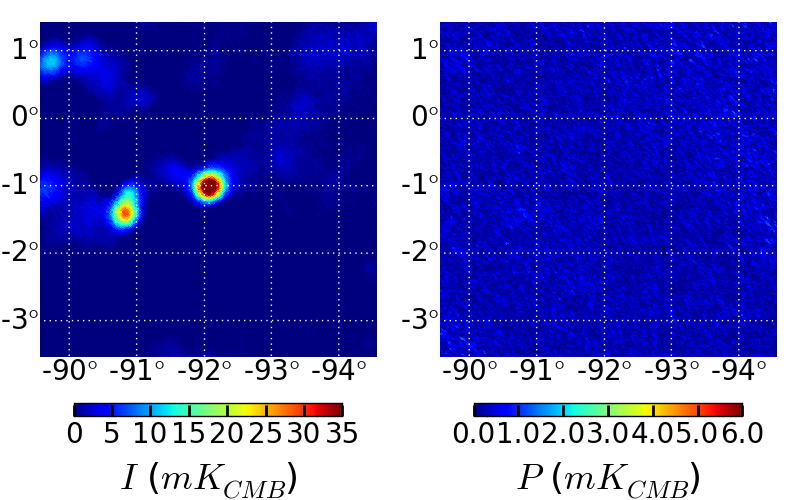}} % TP
\subfigure[``Non-Linear"]{\includegraphics[width=0.38\columnwidth]{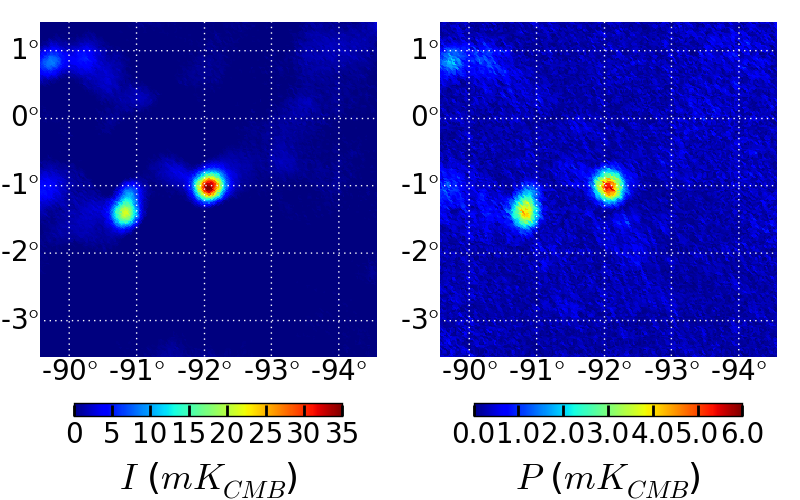}} % TPNL
\subfigure[``ICP removed"]{\includegraphics[width=0.38\columnwidth]{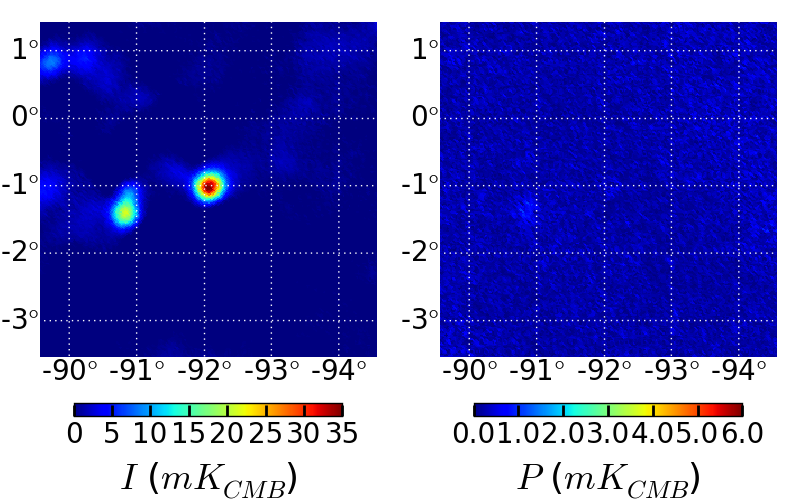}} % TPNL IP removed
}
\caption{ \footnotesize
Simulations of RCW38 maps in intensity and polarization power $P$ for three simulated datasets: ``reference" (left), non-linear (middle), ``ICP removed" (right). 
The maps shown in Galactic coordinates co-add 216 detectors. The polarization orientation is reconstructed in the instrument frame. }
\label{figure: simul rcw38}
\end{figure}

\begin{figure}[h!]
\centering     %%% not \center
\makebox[\textwidth][c]{
\subfigure[``Reference"]{\includegraphics[width=0.36\columnwidth]{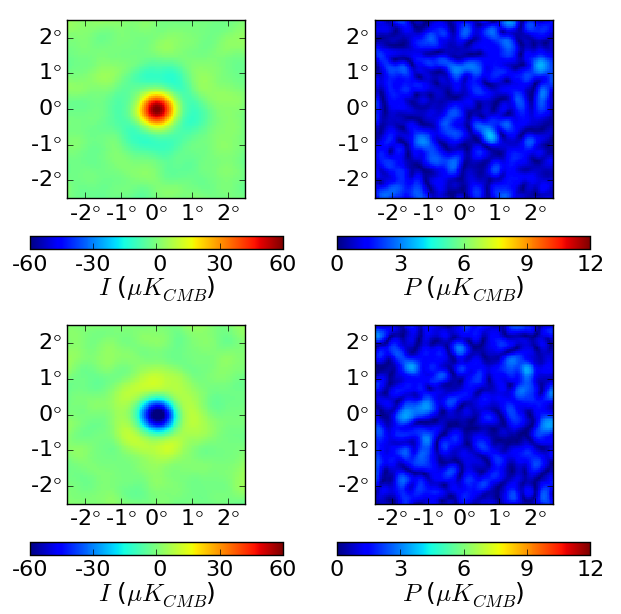}} % TP
\subfigure[``Non-Linear"]{\includegraphics[width=0.36\columnwidth]{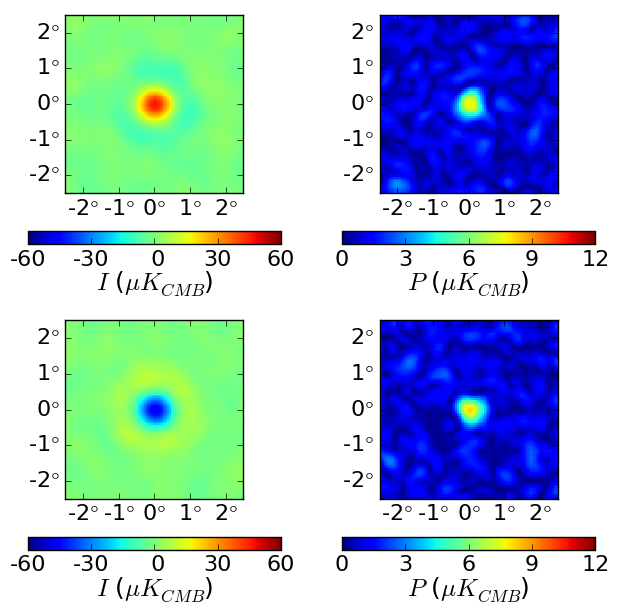}} % TPNL
\subfigure[``\ac{ICP} removed"]{\includegraphics[width=0.36\columnwidth]{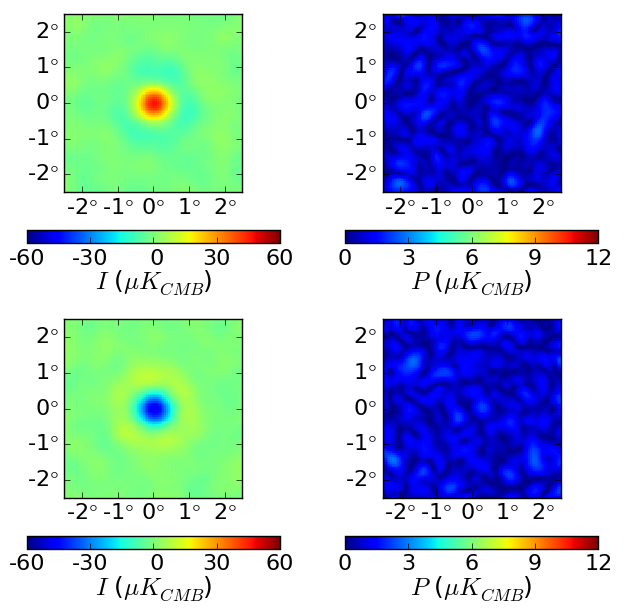}} % TPNL IP removed
}
\caption{ \footnotesize
Simulations of CMB stacked spots in intensity and polarization power $P$ for three simulated datasets: ``reference" (left), ``non-linear" (middle), ``ICP removed" (right). 
%The hot (cold) spots co-add 2122 (1997) extremas. 
The polarization orientation is reconstructed in the instrument frame.
95\% of the 17\% \ac{ICP} is removed in the cleaned dataset (right).
}
\label{figure: simul cmb}
\end{figure}

\begin{figure}[h!]
\centering     %%% not \center
\makebox[\textwidth][c]{
\subfigure[``Reference"]{\includegraphics[width=0.36\columnwidth]{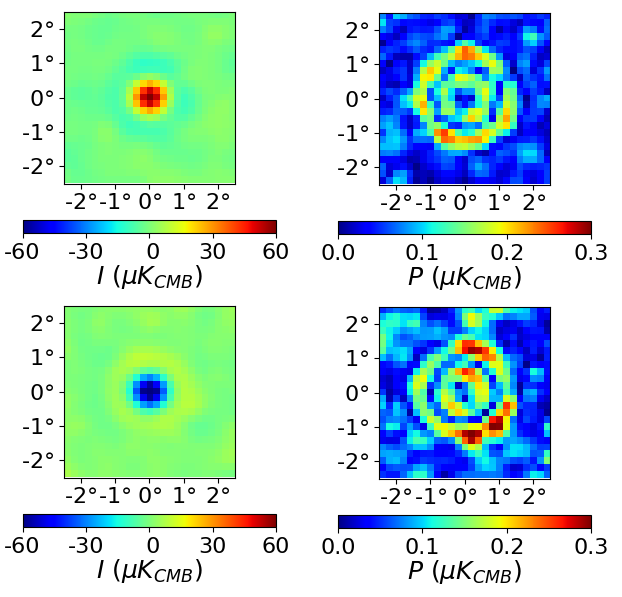}} % TP
\subfigure[``Non-Linear"]{\includegraphics[width=0.36\columnwidth]{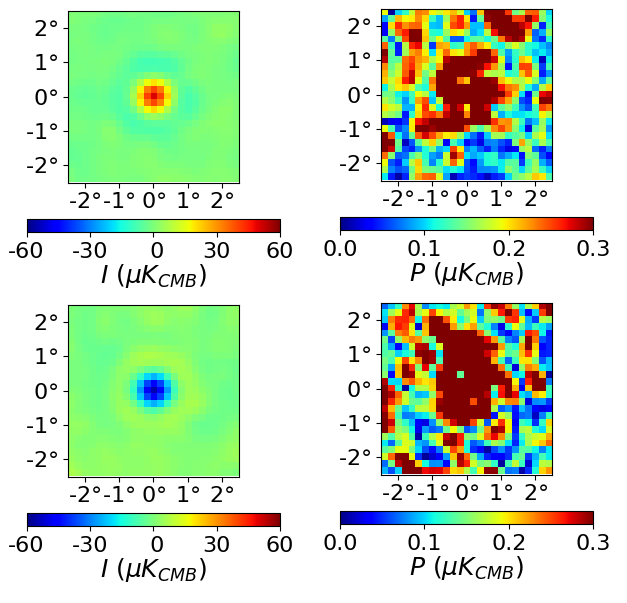}} % TPNL
\subfigure[``\ac{ICP} removed"]{\includegraphics[width=0.36\columnwidth]{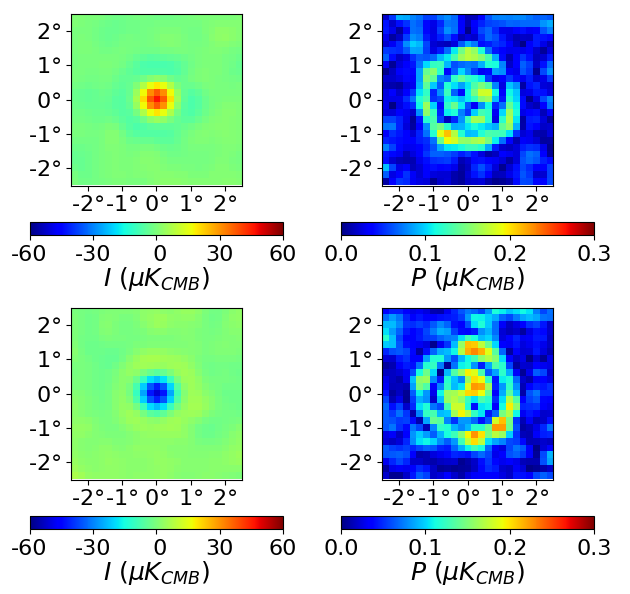}} % TPNL IP removed
}
\caption{ \footnotesize
Simulations of CMB stacked spots in intensity and polarization power $P$ for three simulated datasets: ``reference" (left), ``non-linear" (middle), ``ICP removed" (right). 
The polarization orientation is reconstructed in the sky frame.
Note that the colorscale in polarization is different from the previous stacked spots presented, and that these simulations are noiseless. 
%The hot (cold) spots co-add 2122 (1997) extremas. 
These noiseless maps were produced early in the analysis and have a larger pixel size than other stacked spots presented in the paper. 
}
\label{figure: simul cmb sky frame}
\end{figure}

We measure the coupling parameters of each detector in the simulated non-linear dataset using the map-based method with the Galaxy as a source described in Section~\ref{section: leakage params characterization}. 
With the measured coupling parameters we subtract the ICP from the time-streams using Equation~\ref{eq: cleaned timestreams} to produce the ICP removed dataset, 
and finally we make maps of the cleaned time-stream. 
Note that the removal method only removes ICP, it doesn't correct for other non-linear effects such as the compression of $I^{sky}$ and $P^{sky}$ described in Section~\ref{subsection: detector nl icp}.
We present in Figures~\ref{figure: simul rcw38}, \ref{figure: simul cmb} and \ref{figure: simul cmb sky frame} maps of RCW38 and the stacked CMB spots comparing the three simulated datasets.
Table~\ref{table: simul I to P coeffs} gives quantitative correlations between $I$ and $P$ for the simulated datasets. 
To calculate how much ICP is removed we take the ratio of the difference of polarized power in the non-linear and ICP removed dataset with the polarized power in the non-linear dataset.

\paragraph{RCW38}
Figure~\ref{figure: simul rcw38} shows the RCW38 maps in intensity $I$ and polarization power $P$ for each of the three simulated datasets. 
A 13\% coupling fraction is apparent from the simulated non-linear dataset in the middle panel, 
which is consistent with the 12\% coupling measured in EBEX data (see linear slope in Table~\ref{table: I to P coeffs}). 
The removal of the coupling is evident in the ICP removed dataset (right panel): 98\% of the \ac{ICP}  has been removed. 

%%%%%%%%%%%%%%% CORRELATION TABLE %%%%%%%%%%%%%%%%%%
\begin{table}[h!]
\begin{center}       
\makebox[0.8\columnwidth][c]{
\begin{tabular}{l|ll|ll}
                                           & \multicolumn{2}{c}{RCW 38} & \multicolumn{2}{c}{CMB Stacked Spots}  \\
                                           \hline
                                           & Correlation  & Linear     & Correlation      & Linear  \\
                                           & Coefficient  & Slope (\%) & Coefficient      & Slope (\%)       \\
                                           \hline
Simulation ``reference"                    &    0.0      & 0    &   0.2     & 1 \\
Simulation ``non-linear"                   &    0.9      & 13   &   1.0     & 17    \\
Simulation  ``ICP removed"                 &    0.1      & 0    &   0.3     & 1    \\

        \hline
\end{tabular}
}
\end{center}
%\normalsize
\footnotesize
\caption{\footnotesize Pearson correlation and linear slope (corresponding to an average of the coupling fraction $A'_4$ across detectors) between $I$ and $P$ using RCW38 and stacked CMB spots. For RCW38, only pixels with $I$ greater than 9~\si{\milli\kelvin} are used for calculations. For CMB, only pixels with $I$ greater than 10~\si{\micro\kelvin} are used for calculations.}
\label{table: simul I to P coeffs}
\end{table}
%%%%%%%%%%%%%%%%%%%%% END TABLE %%%%%%%%%%%%%%%%%%%%%%%%%%%%%%

%%%%%%%%%% RCW38 maps -- one column, one line format (subfigure) %%%%%%%%%%%%%%%%%%%%%%%%%%%%%
%%%%%%%%%%%%%% end RCW 38 maps %%%%%%%%%%%%%%%%%%%%

\paragraph{CMB Stacked Spots in the Instrument Frame} 
Figure~\ref{figure: simul cmb} shows the CMB stacked spots constructed from the three simulated datasets. The polarization orientation is co-added in the instrument frame. 
The non-linear time-streams (middle) produce a coupling fraction of 17\% that is visible in the center of the $P$ stacked spots. 
%The correlation and linear slope between $I$ and $P$ for the three datasets are shown in Table~\ref{table: simul I to P coeffs}. 
The coupling coefficients produced by the non-linear simulation are in agreement with those measured in EBEX data (see linear slope in Table~\ref{table: I to P coeffs}). 
In the \ac{ICP} removed dataset, 95\% of the ICP has been removed. 

\paragraph{CMB Stacked Spots in the Sky Frame} 
We plot in Figure~\ref{figure: simul cmb sky frame} the stacked spots in the sky frame for the three simulated datasets. 
When stacking CMB spots in the sky frame, E-modes are apparent as rings of polarization power surrounding the cold and hot $I$ spots.
For this analysis we used noiseless simulations because adding the EBEX2013 level of noise would have entirely obscured the polarization structure apparent in Figure~\ref{figure: simul cmb sky frame} (a).
The 17\% \ac{ICP} completely obscures the E-modes in the non-linear dataset. 
In the ICP removed dataset, 99\% of the ICP is removed and the standard deviation between the input and recovered E-modes is 0.01~uK. 

%%%%%%%%%% CMB maps -- ONE column format %%%%%%%%%%%%%%%%%%%%%%
%%%%%%%%%%%%%% CMB maps %%%%%%%%%%%%%%%%%%%%

%%%%%%%%%% CMB maps -- sky frame -- one row format %%%%%%%%%%%%%%%%%%%%%%
%%%%%%%%%%%%%% CMB maps %%%%%%%%%%%%%%%%%%%%

\clearpage

\subsection{EBEX2013}
\begin{figure}[h!]
\centering     %%% not \center
\makebox[\textwidth][c]{
\subfigure[150 GHz]{\includegraphics[width=0.5\columnwidth]{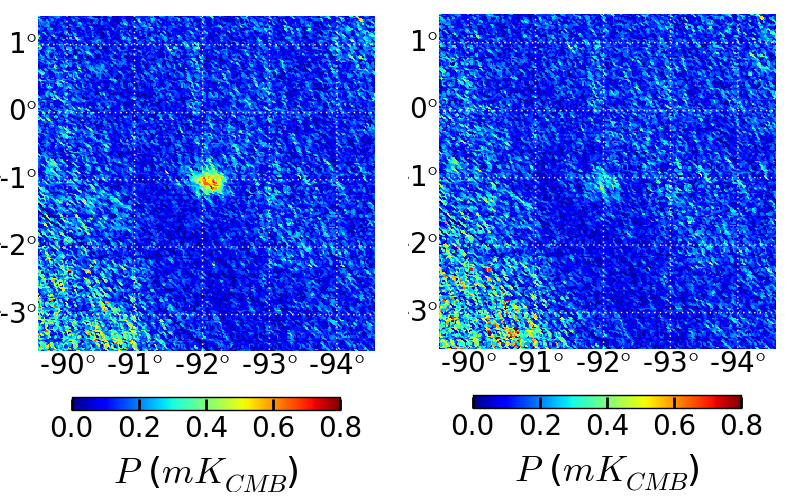}} % 150
\subfigure[250 GHz]{\includegraphics[width=0.5\columnwidth]{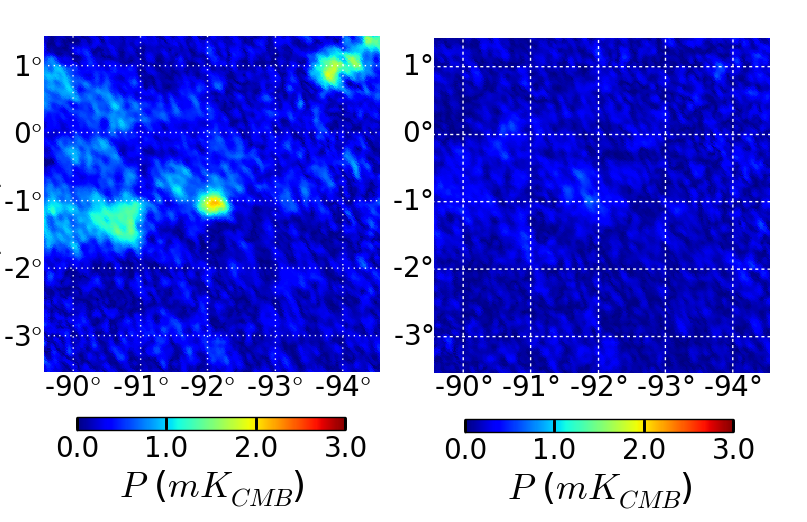}} % 250
}
\caption{ \footnotesize
Comparison of RCW38 maps in polarization power $P$ before (left) and after (right) \ac{ICP} removal for 150~GHz (Figure (a)) and 250~GHz (Figure (b)) EBEX detectors. 
The polarization orientation is reconstructed in the instrument frame.}
\label{figure: rcw38 before after}
\end{figure}

\begin{figure}[h!]
\centering     %%% not \center
\makebox[\textwidth][c]{
\subfigure[150 GHz]{\includegraphics[width=0.5\columnwidth]{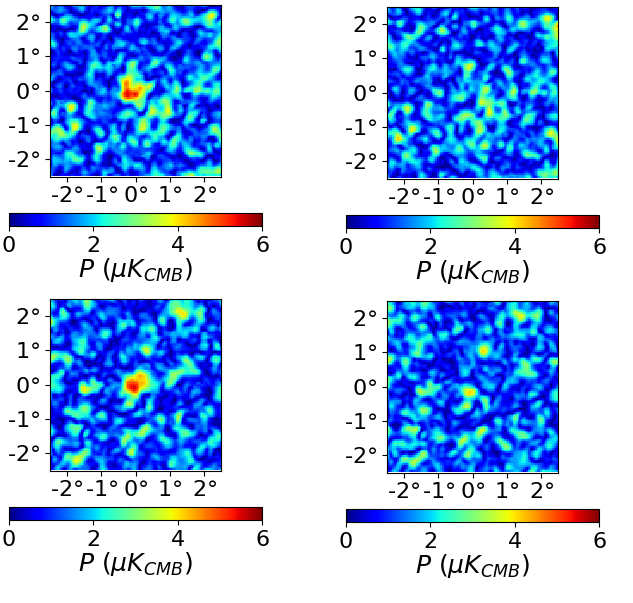}} % 150
\subfigure[250 GHz]{\includegraphics[width=0.5\columnwidth]{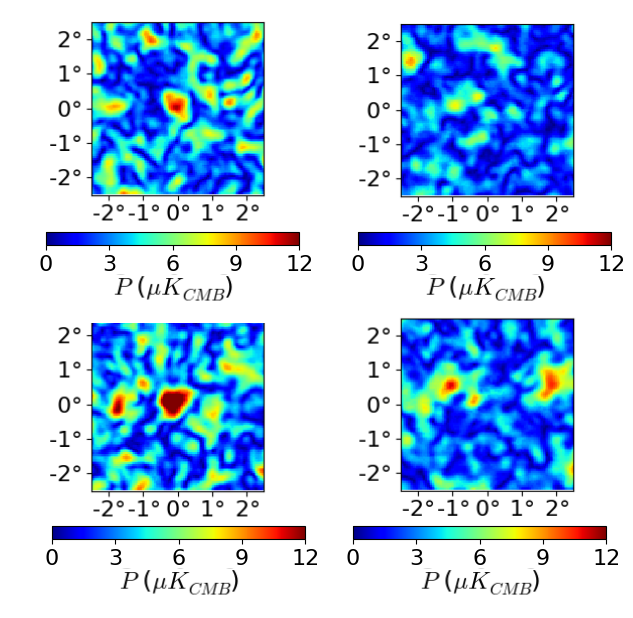}} % 250
}
\caption{ \footnotesize
Comparison of CMB stacked spots in polarization power $P$ before and after \ac{ICP} removal for 150 GHz (left four) and 250 GHz (right four) EBEX detectors. 
Within each frequency band, the maps are shown before (left) and after (right) excess polarization removal. Hot and cold spot are shown on the top and bottom panels, respectively.
The polarization orientation is reconstructed in the instrument frame.}
\label{figure: cmb before after}
\end{figure}

The \ac{ICP} removal procedure is applied to EBEX2013 data, for both 150 and 250 GHz detectors. The Galaxy is used to measure coupling coefficients, which are then used to produce 
time-streams with the ICP removed. 
We present RCW38 maps generated before and after removal in Figure~\ref{figure: rcw38 before after}, and CMB stacked spots in Figure~\ref{figure: cmb before after}. 
Table~\ref{table: I to P coeffs after removal} summarizes the \ac{ICP} Pearson correlation and coupling coefficients in each frequency band before and after removal of the excess polarization. For RCW38, 67\% (98\%)  of the ICP is removed at 150 (250) GHz. In the stacked spots, 81\% (92\%)  of the excess polarization is removed.

%%%%%%%%%%%%%%% CORRELATION TABLE %%%%%%%%%%%%%%%%%%
\begin{table}[h!]
\footnotesize
\begin{center}       
\makebox[0.8\columnwidth][c]{
\begin{tabular}{l|ll|ll}
                                             & \multicolumn{2}{c}{RCW 38} & \multicolumn{2}{c}{CMB Stacked Spots}  \\
                                             \hline
                                             & Correlation  & Linear     & Correlation  & Linear  \\
                                             & Coefficient  & Slope (\%) & Coefficient  & Slope (\%) \\
                                             \hline
EBEX 150 GHz                                 &   0.8   & 11    &   0.8   & 8    \\ 
EBEX 150 GHz with \ac{ICP}  removed          &   0.4   & 3     &   0.2   & 2    \\
EBEX 250 GHz                                 &   0.8   & 12    &   0.6   & 16  \\
EBEX 250 GHz with \ac{ICP}  removed          &   0.5   & 0     &   0.1   & 1   \\
        \hline
\end{tabular}
}
\end{center}
\caption{\footnotesize Pearson correlation and linear slope (corresponding to $A'_4$ averaged over detectors) between $I$ and $P$ after the excess polarization removal.
For ease of comparison, the pre-removal numbers are copied over from Table~\ref{table: I to P coeffs}.} 
\label{table: I to P coeffs after removal}
\end{table}
%%%%%%%%%%%%%%%%%%%%% END TABLE %%%%%%%%%%%%%%%%%%%%%%%%%%%%%%

%%%%%%%%%% RCW38 maps -- real data -- P before/after IP removal  %%%%%%%%%%%%%%%%%%%%%%%%%%%%%
%%%%%%%%%%%%%% end RCW 38 maps %%%%%%%%%%%%%%%%%%%%

%%%%%%%%%% CMB maps -- real data -- P before/after IP removal  %%%%%%%%%%%%%%%%%%%%%%%%%%%%%
%%%%%%%%%%%%%% end RCW 38 maps %%%%%%%%%%%%%%%%%%%%

\subsection{Discussion and Summary}

Continuously rotating HWP are increasingly used in CMB instruments because they reduce \ac{ICP} originating from detector differencing
and because they mitigate low frequency noise enabling observations on large angular scales, which are otherwise limited by atmospheric 
turbulence.  Considering ICP alone, the HWP should be the first element in the optical path, in order to modulate only incident polarized sky 
signals. With the EBEX instrument, which had a 1.5~m entrance aperture, it was not practical for the HWP to be the first element in the optical path. 
We placed it behind the field lens, heat-sunk to a temperature of 4~K, to reduce optical load on the detectors. We anticipated $\ac{ICP}^{IP}$
of up to 2.7\%. The data showed an \ac{ICP} larger than 10\%. We found that the relatively large HWPSS induced non-linear 
detector response, which in turn caused significant conversion of intensity signals to polarization, \ac{ICP}$^{NL}$. 
%An earlier publication by a ground-based instrument showed that such non-linear response also induced low frequency noise in the demodulated polarization bandwidth (GIVE REFERENCE). 

We developed and applied an ICP removal method to the EBEX2013 data, using the Galaxy to measure 
coupling parameters and assessing the quality of the removal on RCW38 maps and CMB stacked spots. 
We showed that for the  EBEX2013 data 81 (92) \% of the ICP was removed from the CMB at 150 (250) GHz using this technique.
The removal of the ICP performs better at 250 GHz compared to 150 GHz. We think this is due to the 
150 GHz detectors having an elliptical beam that is not taken into account during calibration or when using a reference map to measure the 
coupling parameters and subtract the excess polarization (only a symmetrical fit to the beam is used). This would also explain why the ICP 
removal works better on the CMB stacked spots smoothed to 0.5\si{\degree} compared to RCW38 maps which vary on smaller scales. 
We note that the CMB stacked spots are a good test of the quality of the removal as it uses a separate dataset (CMB) than is used to 
compute the coupling parameters (Galaxy). 

%For future CMB experiments, the expected coupling coefficients are of order percent or sub-percent, and this method can be used to remove the ICP and recover the CMB polarization.

The method we presented removes ICP regardless of its source. However, if the detectors have a non-linear response, other effects 
are present in addition to ICP$^{NL}$, such as attenuation of $I^{sky}$ and $P^{sky}$, as demonstrated in the simulations by 
comparing $I$ in the reference and non-linear plots in Figure~\ref{figure: simul rcw38} and \ref{figure: simul cmb}. These effects 
are not corrected by the ICP removal method. Our method, as is the method presented by \cite{polarbear_hwp_2017}, does not correct the non-linearity 
of the detectors and does not correct the distortion induced due to the non-linearity upon incident sky $Q$ and $U$ Stokes signals. The 
level of these distortion needs to be assessed separately, particularly for experiments targeting higher precision polarimetry. Alternatively, 
non-linearity should be avoided by reducing the range of incoming signals (in particular the HWPSS), or by using detectors with operating parameters that ensure linear response 
over a larger dynamic range of incident signals.

\section*{Acknowledgement}
\acknowledgments
Didier acknowledges a NASA NESSF fellowship NNX11AL15H.
Support for the development and flight of the EBEX instrument was provided by
NASA grants NNX12AD50G, NNX13AE49G,
NNX08AG40G, and NNG05GE62G, and by NSF grants AST-0705134 and ANT-0944513.
We acknowledge support from the Italian INFN INDARK Initiative.
Ade and Tucker acknowledge the
Science \& Technology Facilities Council for its continued support of the
underpinning technology for filter and waveplate development.
We also acknowledge support by the Canada Space Agency, the Canada
Research Chairs Program, the Natural Sciences and
Engineering Research Council of Canada, the Canadian Institute for Advanced Research,
the Minnesota Supercomputing Institute, the National Energy Research
Scientific Computing Center, the Minnesota and Rhode Island
Space Grant Consortia, our collaborating institutions, and Sigma Xi the
Scientific Research Society.
Baccigalupi acknowledges support from the RADIOFOREGROUNDS
grant of the European Union's Horizon 2020
research and innovation program (COMPET-05-2015, grant agreement number 687312)
and the INDARK INFN Initiative.

Reichborn-Kjennerud acknowledges an NSF Post-Doctoral Fellowship AST-1102774,
and a NASA Graduate Student Research Fellowship. Raach and Zilic
acknowledge support by the Minnesota Space Grant Consortium.
Helson acknowledges NASA NSTRF fellowship NNX11AN35H.
We very much thank Danny Ball and his colleagues at the Columbia
Scientific Balloon Facility for their dedicated support of the EBEX program.
%We are grateful for contributions to the fabrication of optical elements by Enzo Pascale and
%Lorenzo Moncelsi. 

%% file: tex/acronyms.tex
\begin{acronym}
    %A
    \acro{ACS}{attitude control system}
    \acro{ADC}{analog-to-digital converter}
    \acrodefplural{ADC}[ADCs]{analog-to-digital converters}
    \acro{ADS}{attitude determination software}
    \acro{AHWP}{achromatic HWP}
    \acro{AMC}{Advanced Motion Controls}
    \acro{ARC}{anti-reflection coating}
    \acro{ATA}{Advanced Technology Attachment}
    %B
    \acrodefplural{BRC}[BRCs]{bolometer readout crates}
    \acro{BRC}{bolometer readout crate}
    \acro{BLAST}{Balloon-borne Large-Aperture Submillimeter Telescope}
    %C
    \acro{CAN bus}{Controller Area Network bus}
    \acro{CMB}{cosmic microwave background}
    \acro{CMM}{coordinate measurement machine}
    \acro{CSBF}{Columbia Scientific Balloon Facility}
    \acro{CCD}{charge coupled device}
    %D
    \acro{DAC}{digital-to-analog converter}
    \acrodefplural{DAC}[DACs]{digital-to-analog converters}
    \acro{DASI}{Degree~Angular~Scale~Interferometer}
    \acro{dGPS}{differential global positioning system}
    \acro{DfMUX}{digital~frequency~domain~multiplexer}
    \acro{DLFOV}{diffraction limited field of view}
    \acro{DSP}{digital signal processing}
    %E
    \acro{EBEX}{E~and~B~Experiment}
    \acro{EBEX2013}{}
    \acro{ELIS}{EBEX low inductance striplines}
    \acro{EP1}{EBEX Paper 1}
    \acro{EP2}{EBEX Paper 2}
    \acro{EP3}{EBEX Paper 3}
    \acro{ETC}{EBEX test cryostat}
    %F
    \acro{FDM}{frequency domain multiplexing}
    \acro{FPGA}{field programmable gate array}
    \acro{FCP}{flight control program}
    \acro{FOV}{field of view}
    \acro{FWHM}{full width half maximum}
    %G
    \acro{GPS}{global positioning system}
    %H
    \acro{HPE}{high-pass edge}
    \acro{HWP}{half-wave plate}
    %I
    \acro{IA}{integrated attitude}
    \acro{ICP}{intensity-coupled-polarization}
    \acro{IP}{instrumental polarization} 
    %J
    \acro{JSON}{JavaScript Object Notation}
    %L
    \acro{LDB}{long duration balloon}
    \acro{LED}{light emitting diode}
    \acro{LCS}{liquid cooling system}
    \acro{LC}{inductor and capacitor}
    \acro{LPE}{low-pass edge}
    %M
    \acro{MLR}{multilayer reflective}
    \acro{MAXIMA}{Millimeter~Anisotropy~eXperiment~IMaging~Array}
    %N
    \acro{NASA}{National Aeronautics and Space Administration}
    \acrodefplural{NASA}[NASA's]{National Aeronautics and Space Administration's}
    \acro{NDF}{neutral density filter}
    %P
    \acro{PCB}{printed circuit board}
    \acro{PE}{polyethylene}
    \acro{PTFE}{polytetrafluoroethylene}
    \acro{PME}{polarization modulation efficiency}
    \acro{PSF}{point spread function}
    \acro{PV}{pressure vessel}
    \acro{PWM}{pulse-width modulated}
    \acro{PSD}{power spectral density}
    %R
    \acro{RMS}{root mean square}
    %S
    \acro{SLR}{single layer reflective}
    \acro{SMB}{superconducting magnetic bearing}
    \acro{SQUID}{superconducting quantum interference device}
    \acro{SQL}{structured query language}
    \acro{STARS}{Star Tracking Attitude Reconstruction Software}
    %T
    \acro{TES}{transition edge sensor}
    \acro{TDRSS}{tracking and data relay satellites}
   \acro{TM}{transformation matrix}
    % U
   \acro{UTC}{Coordinated Universal Time}
    % W
   \acro{WMAP}{Wilkinson Microwave Anisotropy Probe}

\end{acronym}

%% file: tex/appendix_coordinates.tex
\section{Coordinates}
\label{appendix: coordinates}

Throughout the paper we alternate between reconstructing the polarization in a frame rotation fixed with the Galactic coordinate system and a frame rotation fixed with the instrument. This is useful to separate polarization originating from the sky against polarization originating from the instrument, as each adds up coherently only in their respective frame orientations. When pointing in a given direction, the two frames are rotated from each other by the instrument Galactic roll angle $\psi_t$. The $\Phi_t$ offset angle from Equation~\ref{eq: basic mueller matrix} can be broken up into:

%when reconstructing $Q^{sky}$ and $U^{sky}$ in the celestial reference frame, $\Phi_t$ depends on the instrument Galactic roll angle $\psi_t$ encoding the rotation between the instrument and the celestial frames: 

\small
\begin{align}
\Phi_t = \Phi' + 2 \psi_t
\end{align}
\normalsize

\noindent where $\Phi'$ is now constant. For simplicity we do not write out $\Phi'$ or the modulation efficiency $\epsilon$ in the paper. For the celestial reference frame we adopt the \ac{WMAP} conventions  \citep{komatsu11}: the polarization that is parallel to the Galactic meridian is $Q$~$>$~0 and $U$~=~0, and the polarization that is rotated by 45\si{\degree} from east to west (clockwise, as seen by an observer on Earth looking up at the sky) has $Q$ = 0 and $U$ $>$ 0. In the instrument frame, positive $Q$ corresponds to linear polarization along the x-axis and positive $U$ corresponds to polarization 45\si{\degree} between the +x and +y directions. The axes are labelled in Figure~\ref{fig: ray_tracing}.

%% file: tex/appendix_dielectric.tex
\section{Effect of a Di-attenuator on Unpolarized Light}
\label{appendix: dielectric}

The field lens acts as a di-attenuator and polarizes light because of differential transmittance between in plane and out of plane incidence. The Mueller matrix of a di-attenuator with in-plane direction forming an angle $\delta$ with the x-axis is:

%\footnotesize
\begin{align}
G(\delta) =\frac{1}{2} \, R(-\delta) \left(
    \begin{array}{cccc}
    \eta & \varepsilon & 0 & 0 \nonumber\\
    \varepsilon & \eta & 0 & 0 \nonumber\\
    0 & 0 & \sqrt{\eta^2-\varepsilon^2} & 0 \nonumber\\
    0 & 0 & 0 & \sqrt{\eta^2-\varepsilon^2}  
    \end{array}
    \right) R(\delta)
\end{align}
\normalsize

\noindent where $R(\delta)$ is the Mueller rotation matrix, $\eta$~$\sim$~$2$ is the sum of the transmittances along the two perpendicular axis and $\varepsilon$~$\sim$~$0$ is the difference between the transmittances of the two axis. The amount of \ac{IP} is characterized by $\varepsilon$, that we name $\varepsilon^{IP}$ in the text. Note that $\varepsilon^{IP}$ increases as the angle of incident light increases, producing more \ac{IP} at the edge of the field lens than in the center. To calculate the effect of the field lens on incoming unpolarized light $I$, the instrument Mueller matrix is modified to include $G(\delta)$:

\begin{equation}
M_{instr} = M_{lp} M_{hwp}(\gamma_t) \, G(\delta) \label{eq: IP mueller matrix}
\end{equation}

\noindent Using Equations~\ref{eq: IP mueller matrix} and \ref{eq: basic detector response}, this results in the following detector time-stream:

\begin{align}
\vec{S}^{IP}_t &= M_{instr}  \label{equation: detector ip}
    \left(
    \begin{array}{c}
    I^{instr}  \nonumber\\
    0 \nonumber\\
    0 \nonumber\\
    0   
    \end{array}
    \right)  \\
D^{IP}_t &= I^{IP}_t \nonumber \\
    &= \frac{1}{2} I^{instr} \, ( \eta + \varepsilon^{IP} \cos(4\gamma_t - 2 \delta))       
%          &= \frac{1}{2}  I^{in} + {\color{magenta} \frac{1}{2}  (\eta-1) I^{in} +  \frac{1}{2} I^{in} \varepsilon \cos(4\psi - 2 \gamma))}
\end{align}

%% file: tex/appendix_nl_response.tex
\section{Non-Linear Response of a TES bolometer}
\label{appendix: detector non-linearity}

In the EBEX detector readout \citep{Aubin_thesis, MacDermid_thesis}, the change in current $i$ coming from a change in power $\delta P$ incident on the detector can be expanded into a series about the equilibrium point $i_0$:

\begin{align}
i(\delta P) - i_0 &=  \diff{i}{P} \delta P + \frac{1}{2} \diff{^2i}{P^2} \delta P^2 + ... \\
                &= a \, \delta P +  b \, \delta P^2 + ...
\end{align}

\noindent We limit our non-linearity model to second order terms and ignore time-constant effects. 
Let us show that $a$ and $b$ have opposite signs, which will justify our subsequent choice of non-linear function. 
The current $i$ is a function of the bias voltage $V$ and the detector resistance $R(T)$:

\begin{equation}
i = \frac{V}{R(T)}
\end{equation}

\noindent such that the current $i$ can be expressed as:

\begin{align}
i(\delta P) - i_0 &= -\frac{V}{R_0^2} \diff{R}{T} \diff{T}{P} \, \delta P + \big(2\frac{V}{R_0^3} \Big[\diff{R}{T}\Big]^2 - \frac{V}{R_0^2} \diff{^2R}{T^2} \big) \Big[\diff{T}{P}\Big]^2\, \delta P^2 \\
                &= a \, \delta P +  b \, \delta P^2
\end{align}

\noindent where $R_0$ is the detector resistance at the equilibrium point. We assume the thermal response to incoming power is linear. $a$ is negative because $R$ increases with increasing temperature $T$, and $b$ is positive because in TES detectors, $\diff{^2R}{T^2} $ is negative high in the transition (during the EBEX flight, $R_0$ was at 65\% to 85\% of its over-biased resistance $R_N$). Dividing by the responsivity $a$, we can write the non-linear detector response as:

\begin{equation}
f^{NL}(\delta P) = \delta P - K \delta P^2
\end{equation}

\noindent where $K > 0$.